\documentclass[a4paper,fleqn,usenatbib]{mnras}

\usepackage{threeparttable}

\usepackage[T1]{fontenc}
\usepackage{ae,aecompl}
\usepackage{comment}

\usepackage{graphicx}   
\usepackage{amsmath}    
\usepackage{amssymb}    
\usepackage{mathabx}    

\title[Magellan/IMACS optical transmission spectrum of WASP-19b]{ACCESS: A featureless optical transmission spectrum for WASP-19b from Magellan/IMACS}

\author[Espinoza et al.]
{N\'estor Espinoza$^{1,2,3,a,b}$, 
Benjamin V.\ Rackham$^{4,5,c}$, 
Andr\'es Jord\'an$^{2,3,1,5}$,
D\'aniel Apai$^{4,5,6,1}$,
\newauthor
Mercedes L\'opez-Morales$^{7,5}$,
David J. Osip$^8$,
Simon L. Grimm$^9$,
Jens Hoeijmakers$^{10,9}$
\newauthor
Paul A. Wilson$^{11}$,
Alex Bixel$^{4,5}$,
Chima McGruder$^{7}$,
Florian Rodler$^{12}$,
Ian Weaver$^{7}$, 
\newauthor
Nikole K. Lewis$^{13}$
Jonathan J. Fortney$^{14}$
Jonathan Fraine$^{13}$
\\
$^{1}$ Max-Planck-Institut f\"ur Astronomie, K\"onigstuhl 17, 69117 Heidelberg, Germany.\\
$^{2}$ Instituto de Astrof\'isica, Facultad de F\'isica, Pontificia Universidad Cat\'olica de Chile,\\
Av. Vicu\~na Mackenna 4860, 782-0436 Macul, Santiago, Chile.\\
$^{3}$ Millennium Institute of Astrophysics, Av. Vicu\~na Mackenna 4860, 782-0436 Macul, Santiago, Chile.\\
$^{4}$ Department of Astronomy/Steward Observatory, The University of Arizona, 933 N. Cherry Avenue, Tucson,\\
AZ 85721, USA.\\
$^{5}$ Earths in Other Solar Systems Team, NASA Nexus for Exoplanet System Science.\\
$^{6}$ Lunar and Planetary Laboratory, The University of Arizona, 1640 E. Univ. Blvd. Tucson, AZ 85721,\\
AZ 85721, USA.\\
$^{7}$ Harvard-Smithsonian Center for Astrophysics, 60 Garden Street, Cambridge, MA 01238, USA.\\
$^{8}$ Las Campanas Observatory, Carnegie Institution of Washington, 
Colina el Pino, Casilla 601 La Serena, Chile.\\
$^{9}$ University of Bern, Center for Space and Habitability, Gesellschaftsstrasse 6, CH-3012, Bern, Switzerland.\\
$^{10}$ Observatoire de Geneve, Chemin des Maillettes 51, 1290 Versoix, Switzerland.\\
$^{11}$ Leiden Observatory, Leiden University, Postbus 9513, 2300 
RA Leiden, The Netherlands.\\
$^{12}$ European Southern Observatory, Alonso de C\'ordova 3107, Casilla 19001, Santiago, Chile.\\
$^{13}$ Space Telescope Science Institute, 3700 San Martin Drive, Baltimore, MD 21218 USA.\\
$^{14}$ Department of Astronomy and Astrophysics, University of California, Santa Cruz, USA.\\
$^{a}$ Bernoulli Fellow.\\
$^{b}$ Gruber Fellow.\\
$^{c}$ NSF Graduate Research Fellow.
}

\date{Accepted XXX. Received YYY; in original form ZZZ}

\pubyear{2018}

\begin{document}
\label{firstpage}
\pagerange{\pageref{firstpage}--\pageref{lastpage}}
\maketitle

\begin{abstract}
The short period ($0.94$-day) transiting exoplanet WASP-19b is an exceptional target 
for transmission spectroscopy studies, due to its relatively large 
atmospheric scale-height ($\sim 500$ km) and equilibrium temperature ($\sim 2100$ K). 
Here we report on six precise spectroscopic Magellan/IMACS observations, five 
of which target the full optical window from $0.45-0.9\mu$m and one targeting 
the $0.4-0.55\mu$m blue-optical range. Five of these datasets are consistent with 
a transmission spectrum without any significant spectral features, while one shows a significant slope as a function of wavelength, 
which we interpret as arising from photospheric heterogeneities in the star. Coupled 
with HST/WFC3 infrared observations, our optical/near-infrared measurements point to 
the presence of high altitude clouds in WASP-19b's atmosphere in agreement with previous 
studies. Using a semi-analytical retrieval approach, considering both planetary and stellar 
spectral features, we find a water abundance consistent 
with solar for WASP-19b and strong evidence for sub-solar abundances for optical absorbers such as 
TiO and Na; no strong optical slope is detected, which suggests that if hazes are present, 
they are much weaker than previously suggested. In addition, two spot-crossing events are 
observed in our datasets and analyzed, including one of the first unambiguously detected \textit{bright} spot-crossing events on an exoplanet host star.
\end{abstract}

\begin{keywords}
techniques: spectroscopic -- planets and satellites: atmospheres -- 
planets and satellites: individual: WASP-19b --
stars: individual: WASP-19 -- stars: activity -- stars: starspots
\end{keywords}



\section{Introduction}


Transmission spectroscopy, the study of the variation of the planetary radius as a function of wavelength due to different opacity sources in its atmosphere \citep{ss:2000, Brown:2001, hubbard:2001, BSH:2003,fortney:2005}, offers  
one of the most successful approaches to date for detecting atomic and molecular absorption in exoplanet atmospheres.
From detections of water vapor \citep[see, e.g.,][]{Wakeford:2013,Huitson:2013,Fraine:2014,KreidbergW43:2014,Kreidberg:2015,Fischer:2016,Evans:2016,Wakeford:2017}, sodium and potassium \citep{Charbonneau:2002,Redfield:2008,SingK:2011,SingNa:2012,Pont:2013,NikolovHP1:2014,Nikolov:2015,Sing:2015,Nikolov:2016,Fischer:2016, Sing:2016} and many signatures of aerosols, i.e., clouds and/or hazes \citep[see, e.g.,]{Lecav:2008,Jordan:2013,Kreidberg:2014,Nikolov:2015,Sing:2016}, transmission spectroscopy studies are paving the way for us to understand the compositions 
of distant worlds. This exploration is not only interesting in that it reveals 
the diversity of compounds present in these distant worlds and the 
thermochemical processes at play \citep[see, e.g.,][]{madhusudhan:2012,Moses:2013}, but it can also offer a glimpse into the formation mechanisms that generated the atmospheric composition we observe 
today \citep{oberg:2011,Moses:2013,mordasini:2016,Mad:2017,Espinoza:2017}.

Among the exoplanets amenable to atmospheric characterization via transmission spectroscopy, the ultra-short period ($P=0.94$-day) exoplanet WASP-19b \citep{Hebb:2010} orbiting a G8V star is a very interesting one in many ways. It is not only 
inflated \citep[$R_p = 1.410\pm 0.03 R_J$, $M_p=1.139\pm 0.05M_J$][]{Mancini:2013} but also, due to its proximity to its host star, has a high equilibrium temperature ($T = 2,100$ K). This, in turn, gives 
it a large atmospheric scale-height of order $H\sim 510$ km, which should give rise to a signal in transmission between $200-600$ ppm. Indeed, low-resolution transmission spectroscopy studies in the infrared using the Hubble Space Telescope ({\em HST}) have succeeded 
in detecting water vapor absorption in its atmosphere, with amplitude of $300$ ppm \citep{Huitson:2013,Sing:2016,Iyer:2016}. 

What this infrared signature actually means in terms of the actual water abundance 
and what it means to, e.g., the C/O ratio of its atmosphere or its metallicity, however, is still unclear. On one hand, \cite{Huitson:2013} identified the signature as consistent with that from a clear atmosphere, while \cite{Iyer:2016} interpreted the signature as arising from a cloudy atmosphere. Optical signatures could, in principle, help us interpret the observed infrared water signature; however, they have 
proven to be elusive mainly due to the lower precisions in most previous optical works from both space and ground-based observatories compared to those in the infrared from HST \citep[see, e.g.][]{Huitson:2013, Mancini:2013, Sedaghati:2015}. 

Optical absorbers are fundamental to 
interpret infrared water absorption signatures. If no 
absorbers are detected, this would strengthen the case for clouds in the atmosphere, which could lead to a large water abundance in the exoplanet given its large observed water signature. This could hint, in turn, to a low C/O ratio in its atmosphere, 
which would make it consistent with our understanding of solid enrichment in planetary 
envelopes \citep{mordasini:2016,Espinoza:2017}. On the other hand, the actual detection of optical signatures could help us understand many underlying processes in these hot exoplanets, including cloud formation and whether cold traps in the night side could sequester 
Ti or V from the terminator region probed in transmission, preventing the formation 
of TiO/VO expected from equilibrium 
chemistry calculations \citep{Parmentier:2013,Parmentier:2016}. Without the strong optical absorbers of TiO/VO, Na 
and K should be the dominant optical absorbers in a clear atmosphere; detecting any 
of those, thus, would be exciting in terms of understanding the underlying dynamics 
in a hot exoplanet such as WASP-19b. A-priori, however, one would expect the atmosphere to be cloudy, given the ubiquity of clouds in giant exoplanet atmospheres \citep[see, e.g.,][and references therein]{Iyer:2016}. Data, of course, are the ultimate judge that should guide our a-posteriori beliefs.

Motivated by the exciting possibilities in the optical for WASP-19b, we obtained observations for this exoplanet as part of the Arizona-CfA-Cat\'olica-Carnegie Exoplanet Spectroscopy 
Survey (ACCESS). ACCESS is a multi-institutional effort which 
aims at obtaining a large, homogeneous library of optical spectra of exoplanet atmospheres,
providing a complement to longer-wavelength observations and key information necessary to understand 
and interpret observed molecular and/or atomic features. To this end, 
ACCESS is currently using the Inamori-Magellan Areal Camera and Spectrograph \citep[IMACS][]{IMACS:2011} instrument mounted at the Magellan Baade 6.5m Telescope in Las Campanas Observatory in Chile. This 
project has allowed us to study so far the atmospheres of more than a dozen exoplanets including the giant exoplanet 
WASP-6b \citep{Jordan:2013} and the sub-Neptune GJ~1214b \citep{Rackham:2017}.

During the writing of this manuscript, \cite{Sedaghati:2017} presented an optical transmission spectrum of WASP-19b from VLT/FORS2 measurements obtained at three different epochs in three different but overlapping wavelength ranges (4,400--6,100 \r{A}, 5,400--8,460 \r{A}, and 7,400--10,000 \r{A}).
Their combined spectrum shows structures which they interpret as (1) a strong scattering slope, (2) strong TiO signatures and (3) a water signature in the 
near-infrared, all of which point to a clear atmosphere in the optical. 
This finding is in contrast to previous low-resolution 
observations with HST/STIS \citep{Huitson:2013,Sing:2016} showing a relatively 
featureless spectrum, which was interpreted as evidence for clouds and lack of TiO absorption. 
In this work, we present six Magellan/IMACS observations obtained between 2014 and 2017, five of which cover the same wavelength 
range over which most of the strong optical signatures are seen by \cite{Sedaghati:2017} (i.e., from 4,500 to 9,500 \r{A}), and one observation which covers the bluest wavelengths (4,200--5,500 \r{A}), where the strong scattering signature is observed in that work. We do not observe this strong scattering slope 
in our data, nor do we detect the strong TiO absorption feature reported by \cite{Sedaghati:2017}.

This paper is structured as follows. In Section \ref{sec:data} we present our Magellan/IMACS observations of WASP-19b along with the photometric monitoring of the host star available around the 
times of these observations. In Section \ref{sec:analysis} we present 
the analysis of the Magellan/IMACS data, which includes the determination of the transmission spectra for each of the transit observations as well as an analysis of two observed spot-crossing events and a study of the effect of unocculted stellar 
photospheric heterogeneities on our measurements. 
In Section \ref{sec:discussion} we discuss the implications of our findings in light of previous attempts at detecting optical absorbers in the atmosphere of this exoplanet and the {\em HST} infrared observations, and in 
Section \ref{sec:conclusions} we outline our conclusions.

\section{Data}


\label{sec:data}

The data presented in this work considers both spectrophotometric measurements and long-term photometric monitoring of the star. The former, described in detail in Section \ref{sec:imacs}, consists of six transits observed with Magellan/IMACS in 2014, 2015 and 2017, which we use to extract WASP-19b's transmission spectrum in Section 
\ref{sec:analysis}. The long-term photometric monitoring, on the other hand, 
consists of data taken during the 2014 season with the SMARTS 1.3m telescopes at 
Cerro Tololo Inter-American Observatory (CTIO) and data gathered in 2017 by 
the All-Sky Automated Survey for Supernovae \citep[ASAS-SN;][]{ASASSN:2014,ASASSN:2017}. 
These photometric data and their importance to the present work are presented in 
Section \ref{sec:phot-mon}. 

\subsection{Observations with Magellan/IMACS}
\label{sec:imacs}
Transit observations of WASP-19b were carried out with the Inamori-Magellan Areal Camera and Spectrograph \citep[IMACS,][]{IMACS:2011} instrument mounted at the Magellan Baade 6.5m Telescope 
in Las Campanas Observatory (LCO) in Chile in 2014 (March 22 and April 29), 2015 (June 3rd) and 2017 (February 22, 
April 4th and April 12). Hereafter we refer to these datasets by their observation date in the YY/MM/DD format. The $f/2$ camera was used in the multi-object spectrograph configuration, and a mask was cut with $10''$ wide slits at the positions of the target and seven comparison 
stars. The slit size was chosen to be large enough to avoid significant slit losses with typical seeing conditions at LCO. 

One of the key features of Magellan/IMACS is 
its large field of view (FOV) for the $f/2$ camera ($27'\times 27'$), which allows for observations of several 
comparison stars in order to correct for common time-series variations. This is in contrast with other 
instruments, such as VLT/FORS2 
\citep{FORSW19:2015,Sedaghati:2016,Nikolov:2016,Sedaghati:2017,Gibson:2017}, or 
Gemini/GMOS  \citep{Huitson:2017,Stevenson:2014,Gibson:2013b,Gibson:2013a}, whose relatively 
smaller FOVs ($\sim 7'\times 7'$ and $2'\times 2'$, respectively) typically only allow 
the observations of a single comparison star. The availability of multiple comparison stars 
makes it easier to detect and confrim features in the data, including spot-crossing events, which are 
known to occur during WASP-19b transits \citep[see, e.g.,][]{Tregloan-Reed:2013, 
Mancini:2013, Huitson:2013} and which could bias our measurements if not accounted for. A 
list of the comparison stars observed simultaneously with WASP-19 in our study is shown in Table \ref{tab:comps}; the star J09545906-4544371 ($V = 10.3$), however, was not 
used in the final analysis as the observed spectra saturated the detector in the wavelength range of interest. 

The observations during the 2014 and 2015 season were performed using the \texttt{TURBO} 
(30s) readout mode, with the 300+17.5 (300 lines/mm and 17.5 degree blaze angle) 
grism, which provided spectra spanning the 4,100--9,350~\AA\ range and an adequate spectral resolution for 
our purposes ($\sim 1.3$~\AA\ per pixel). The observations obtained during the 2017 season were carried 
out with the \texttt{Fast} (31s) readout mode and a blocking filter (BF) in order to assess if second order 
light could be impacting our observations. The February 22 and April 4th 2017 observations were carried out 
using the same grism as the 2014 and 2015 seasons, but the GG455 blocking filter, which has a sharp cutoff 
at 4,550 \AA\, was used; we refer to this mode as the ``red" setup. For the 17/04/12 observations, we aimed to study the blue part of the transmission spectrum, and thus we 
decided to use the same grism but with the WB3600-5700 blocking filter, which allowed us to block any light 
bluewards of 3,600 \AA\ and redwards of 5,700 \AA. We refer to this mode as the ``blue" setup. All observations were acquired with a 
$2\times2$ binning. The data were reduced with a pipeline already outlined in previous works \citep{Jordan:2013,Rackham:2017}, and which will be fully detailed in a future paper (Espinoza et al., 2018, in prep.).   
 
\begin{table}
    \centering
    \caption{Comparison stars used in this study. First column identifies the 2MASS identifier 
    of the comparison star, second and third its RA and DEC, and fourth column its $V$ magnitude.}
    \label{tab:comps}
    \begin{tabular}{lccc} 
        \hline
        \hline
2MASS ID & RA (J2000.0) & DEC (J2000.0) & $V$\\
        \hline
J09543557-4537090 & 09:54:45.612 & -45:38:03.29 & 12.8\\
J09524470-4540273 & 09:52:44.706 & -45:40:27.37 & 13.6\\
J09540698-4544274 & 09:54:09.354 & -45:45:25.49 & 11.6\\
J09545906-4544371 & 09:54:59.060 & -45:44:37.20 & 10.3\\
J09535442-4540018 & 09:53:54.430 & -45:40:01.83 & 13.3\\
J09531573-4531570 & 09:53:15.734 & -45:31:57.11 & 13.2\\
J09535727-4546424 & 09:53:57.279 & -45:46:42.45 & 12.05\\
        \hline
    \end{tabular}
\end{table}

\subsection{Photometric monitoring}
\label{sec:phot-mon}
WASP-19 is known to display photometric variability at the $5$-$10$ mmag level \citep{Hebb:2010,Huitson:2013}, which is most likely caused by the rotational modulation of starspots \citep{Tregloan-Reed:2013}. 
It has been long recognized that starspots (even if unocculted) can impact transiting exoplanet transmission spectra
by introducing
time- and wavelength dependent biases into the apparent size of the planet \citep[see][]{Sing:2011, Pont:2013, Huitson:2013, Rackham:2017, TLSE:2018}. 
The origin of the spectral contamination is the spectral difference between the disk-averaged spectrum (obtained pre-transit) and the spectrum of the transit chord (the actual light source for the transmission measurement), which has been termed the transit light source effect (TLSE) by \cite{TLSE:2018}. 
Several studies have recognized and dealt with this effect in the past with different 
levels of complexity \citep[see, e.g.,][]{Pont:2008,Sing:2011,Deming:2013}. 
However, the work of \cite{TLSE:2018} points out that the correction of the impact of starspots and faculae, 
which can introduce not only apparent spectral slopes but also false spectral features in transmission, is not straightforward. 
In particular, they demonstrate that the commonly assumed linear relation between stellar photometric variability and starspot/facular areal covering fractions will underestimate the covering fractions in most realistic cases. 
Nevertheless, {\em in principle} if spots/faculae covering fractions and their temperatures are known, one should be able to constrain their impact on the transmission spectrum. 
Such constraints can be informed by photometric monitoring of the exoplanet host star, especially if data were acquired close to the epochs of the transmission spectroscopic observations. 

Considering this, we decided to obtain and/or use photometric observations 
when possible/available. During the 2015 season we were not able to obtain 
photometric monitoring of WASP-19. However, during the 2014 and 2017 seasons, 
photometric data covering the epochs of our Magellan/IMACS observations are 
available. We analyze those datasets in detail in Appendix \ref{sec:phot-mon-an}. Here 
we simply state some key results from that analysis, that we will use in future sections, 
especially in Section \ref{analysis:heter} in which we use the observed level of variability 
to constrain the possible spot covering fractions and thus predict the expected 
level of stellar contamination in WASP-19b's transmission spectrum.

For the 2014 season, WASP-19 was monitored with the SMARTS 1.3m telescope at CTIO through 
a $V$ filter in a continuation of the long-term photometric campaign presented by \citet{Huitson:2013}.
However, strong systematics in the data precluded any 
meaningful photometric variability measurements for WASP-19, despite the fact that we 
used the same observational setup as the long-term photometric 
monitoring presented in \cite{Huitson:2013}. 
We re-analyzed the \cite{Huitson:2013} photometric dataset 
in order to verify that this was not an artifact of our data reduction pipeline and 
recovered the same results for WASP-19 as the ones presented in that work. However, 
the comparison stars all showed similar levels of variability as WASP-19 in that dataset, implying that the observed variability of WASP-19 
in the work of \cite{Huitson:2013} was mostly due to systematic effects and not due 
to intrinsic astrophysical variability (see Section \ref{sec:phot-mon-2014} 
for details). This implies that the stellar activity 
corrections made in that work to the transmission spectrum of WASP-19b are most likely 
not adequate. This is, however, unimportant for the overall shape of the transmission 
spectrum, albeit for a possible offset in transit depth between the optical HST/STIS and 
infrared HST/WFC3 data presented in \cite{Huitson:2013}. This will be important to consider 
in Section \ref{sec:discussion}, when we interpret the HST optical and infrared observations 
in light of our Magellan/IMACS optical data.

For the 2017 season, ASAS-SN V-band photometry was gathered from the ASAS-SN webportal\footnote{\url{https://asas-sn.osu.edu/}}. This data covered all of 
our Magellan/IMACS observations during the 2017 season, allowing the variability 
of WASP-19 during that observing season to be estimated (see Section 
\ref{sec:phot-mon-2017} for details). We observed a
2\% peak-to-peak variability for WASP-19, similar to what has been 
observed in previous works \citep[see, e.g.,][]{Hebb:2010}.

\section{Analysis}
\label{sec:analysis}

In this Section we analyse both our band-integrated (``white light'') and 
wavelength-dependent Magellan/IMACS light curves for WASP-19 and the observed comparison stars, 
extracted from the spectra described in 
Section \ref{sec:data}. 
Our approach is to optimally use the available information in order to account for systematic trends in the data and to analyse the datasets in a uniform way in order to introduce 
the least possible amount of variation between them. 
To this end, we take into account both the comparison stars, which provide essential information 
to correct common atmospheric and instrumental effects present in our light curves, and 
external parameters (such as, e.g., the variation of the full-width at half maximum, FWHM, 
the movement of the spectral trace, etc., during the course of the observations), which can aid in 
correcting systematic effects intrinsic to our WASP-19 light curves. The modelling approach 
is first described in detail in Section \ref{sec:lc-modelling}, and this is later used 
in order to analyse the band-integrated (Section \ref{sec:white-light}) and wavelength-dependent 
(Section \ref{sec:transpec}) light curves. 

\subsection{Light curve modelling}
\label{sec:lc-modelling}


We modeled the flux of WASP-19 as the multiplicative combination of signals and, thus, we work in logarithmic space (base 10) in order to represent components as linearly additive. 
In what follows, however, we transform all products (e.g., plots, precision of our best-fit models) to relative flux space for ease of comparison with other studies. 

We follow the approach of \cite{Jordan:2013} and perform a Principal Component Analysis (PCA) in logarithmic space on our 6 comparison stars in order to optimally use all the information embedded in our comparison stars regarding common systematic effects (i.e., present in every star in the field).
We extract the 6 signals, $S_i(t)$, which allow us to (linearly) reconstruct their observed flux variations. 
As explained in \cite{Jordan:2013}, for each signal $S_i(t)$, an eigenvalue $\lambda_i$ allows us to decide how important -- in terms of information content -- that signal is to reconstruct the original flux variations of the comparison stars (and, we assume, to explain the flux variations in WASP-19's light curve as well). 
In what follows, we assume the indices $i$ of the signals order the signals from the most important (larger eigenvalue, $i=0$ in our case) to the least important (smaller eigenvalue, $i=5$ in our case). 

In order to account for systematic effects which might be unique to WASP-19's light curve (due to, e.g., color differences, detector and/or instrumental artifacts), we also include external parameters that are recorded during our observations as possible linear (in logarithmic space) regressors.
These are standardized by subtracting their means and dividing by their standard deviations. 
In this study, we observe that the rotator angle as a function of time $R(t)$, the wavelength drift of the wavelength solution as a function of time $\Delta \lambda (t)$, the variation of the full-width at half maximum as a function of time $\textnormal{FWHM}(t)$, time $t$, and the squares of the mentioned variables (i.e., $R^2(t)$, $\Delta \lambda^2 (t)$, $\textnormal{FWHM}^2(t)$ and $t^2$) are useful regressors that account for the observed variations in our WASP-19 light curves. 
However, determining which of these regressors are the most important to include in our model for each of our light curves is not trivial. 
The same goes for the number of signals $S_i(t)$. 

In order to incorporate our ignorance regarding a ``best-fit" model in our retrieved transit parameters, we follow the model averaging technique outlined for transmission spectroscopy studies in \cite{ModelAveraging:2014}, which has already been used and tested with both space-based \citep[see, e.g.,][]{Wakeford:2016} and ground-based \citep[see, e.g.,][]{Nikolov:2016} observations. 
We define $k$ models for the base-10 logarithm of WASP-19's flux of the form 
\begin{eqnarray}
\label{eq:themodel}
M_k(t) \sim  C_{k} + \sum_{i=0}^{N_k} \alpha_{i,k} S_{i}(t) + X_{k}(t) + \log_{10} T(t|\mathbf{\theta_t}) + \epsilon_k.
\end{eqnarray}
Here, $C_k$ is a zero point in (log-)flux for the $k$-th model, $N_k$ is the number of signals $S_i(t)$ to be included in the fit weighted by $\alpha_{i,k}$ in the $k$-th model, $X_{k}(t)$ is a linear combination of up to 3 of the 8 regressors mentioned in the previous paragraph\footnote{The decision of up to 3 terms was empirically based on our best signal-to-noise dataset, in which we observed that more terms only slow down the process without additional improvement on the fits.}, $T(t|\mathbf{\theta_t})$ is the transit model with parameters $\mathbf{\theta_t}$, and $\epsilon_k$ is a random variable assumed to be gaussian and uncorrelated and, thus, defines a photometric jitter term $\textnormal{Var}[\epsilon_k] = \sigma^2_k$ empirically determined for each fit. 
This defines a total of 552 models to be tested on our data if all signals $S_i(t)$ are incorporated in our framework (6), and models of up to 3 terms out of 8 possible variables are generated (i.e., $\sum_{n=1}^3 \binom{8}{n}=92$ possible models). 
This number varies slightly for the wavelength-dependent light curves, because of the varying spectral coverage of the comparison stars (i.e., some of them do not cover the full spectral range covered by WASP-19 due to the positioning of the mask slits relative to chip gaps and detector edges). 

In practice, after removing obvious outliers from our light curves using a sigma-clipping procedure, each model is first fit with a Levenberg-Marquardt algorithm using the out-of-transit data in order to estimate the linear parts of the model ($C_k$, the coefficients of the signals $S_i(t)$ and $X_k(t)$). 
Then, the data are detrended with this fit and initial transit parameters $\mathbf{\theta_t}$ are estimated by fitting the non-linear transit model also using a Levenberg-Marquardt algorithm.
The retrieved parameters in this two-step process are then fed to an MCMC algorithm as starting points, and using \texttt{emcee} \citep{emcee:2013}, the posterior distribution of all the parameters in the full model is explored (including now the jitter term, $\sigma^2_k$). 
The median of the posteriors are then defined as the best-fit parameters of the $k$-th model.
The corrected Akaike's Information Criterion \cite[AICc; see][and references therein]{cavanaugh:1997} of each model is then computed, and the final, model-averaged parameters and uncertainties are obtained by weighing the best-fit parameters of each model by their respective AICc values as outlined in \cite{ModelAveraging:2014}. 

As a final note on our modelling, as noted in \cite{EJ:2015}, our limb-darkening assumptions can lead us to retrieve biased transit parameters if not properly accounted for in the transit light curve fitting procedure, especially for very precise transit light curves like those we present in the next sub-section. 
In addition, for active stars such as WASP-19\footnote{Here and in what follows, we refer to ``activity" as stars producing large rotational modulations and/or having evidence of heterogeneities in their photospheres.}, the limb-darkening coefficients can actually be modified due to the non-uniform stellar surface owing to the presence of unocculted spots \citep{Csizmadia:2013}. 
We therefore chose to fit the limb-darkening coefficients in our transit light curve fitting procedure rather than to fix them to pre-computed values. 

To select the optimal limb-darkening law to use in our case, we use the method outlined in \cite{EJ:2016}; 
in brief, we generate light curves with similar geometric and noise properties as the ones 
observed using a non-linear limb-darkening law and then try to retrieve the transit parameters using two-parameter laws and the linear law. 
We then identify the law that retrieves the parameters closest to the input values in a mean-square error sense as the optimal law to use. 
We observed that the square-root law is, in this case, the optimal one for noise levels as low as $\sim 1000$ ppm, and thus we use this limb-darkening law throughout our analysis of the band-averaged (i.e., "white-light") light curves. 
Beyond this noise level, our analysis shows that the linear law is as good as the other tested laws, and so we use it for the wavelength-dependent light curves. 
The \texttt{batman}\footnote{\url{https://www.cfa.harvard.edu/~lkreidberg/batman/}} \citep{batman:2015} package was used to model the transit light curves, and the limb-darkening coefficients were sampled following the uninformative sampling procedure outlined in \cite{kipping:2013}. 
The routines used to model our transit light curves are available at GitHub\footnote{\url{http://www.github.com/nespinoza/exotoolbox}}.

\subsection{``White-light" light curve analysis}
\label{sec:white-light}


The band-integrated (``white-light") light curves were obtained by following the scheme outlined in Section \ref{sec:lc-modelling}. 
In this case, $250$ walkers are used on each of the $k$ models in order to explore the parameter space. A total of $1000$ steps are used, discarding the first $500$ as burn-in.

\begin{figure*}
   \includegraphics[height=0.45\columnwidth]{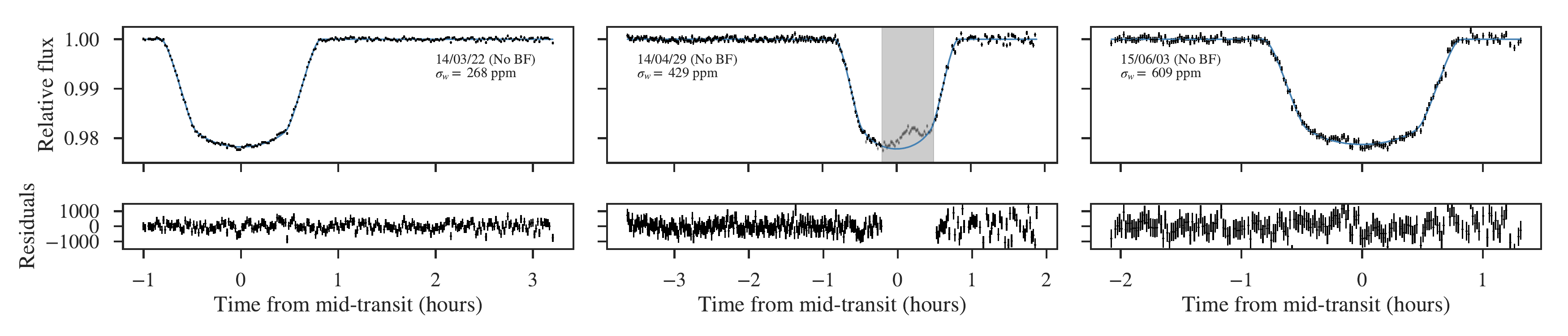}
   \includegraphics[height=0.45\columnwidth]{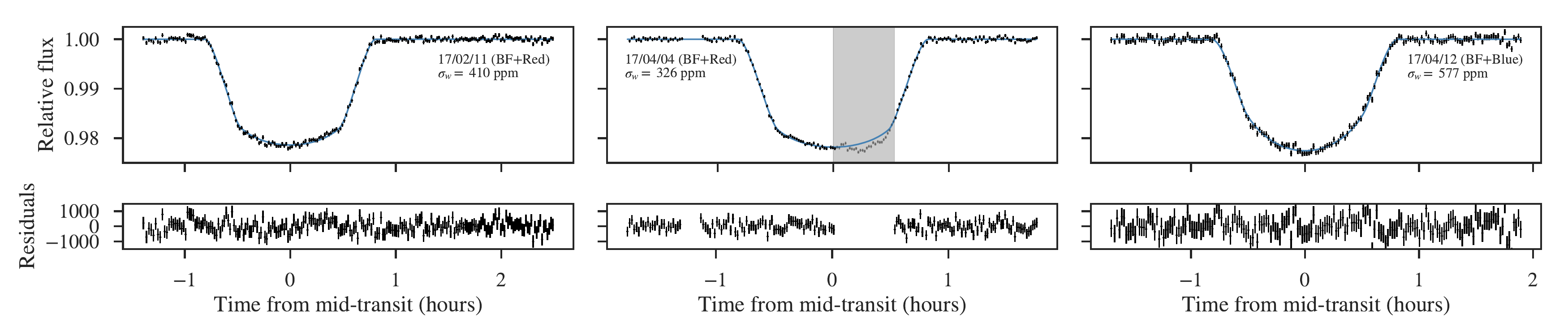}
    \caption{Band-integrated (i.e., "white-light") light curves for our Magellan/IMACS observations with the model-averaged systematics removed (black points with errorbars), along with the corresponding residuals (in parts-per-million, ppm). The best-fit transit light curve given our model averaging procedure is depicted in blue. The value of $\sigma_w$ given in each panel is the model-averaged value. Grey bands denote points left out of this white-light analysis, as they contain evident spot-crossing events.}
    \label{fig:white-transits}
\end{figure*}

In Figure \ref{fig:white-transits}, we present our systematics-removed transit light curves along with the residuals for each model.
The systematics models were obtained by averaging each of the $k$ models fitted to the data following the procedures outlined in Section \ref{sec:lc-modelling}, weighted by their AICc values. 
It is evident that spot-crossing events are present in the 14/04/29 and 17/04/04 light curves and that the features are wavelength-dependent. 
For simplicity, these portions of the light curves were left out of the white-light analysis.
These portions, however, are used and modelled in the analysis of the wavelength-dependent light curves (see next sub-section). 
Following the procedures we outline in Section \ref{sec:spot-crossing}, we find no evidence for spots in the other light curves.

The precision of our white-light light curves is generally 4--5 times the 
photon noise in our datasets, except for the 14/03/22 dataset in which the precision is only 3 times the photon noise. 
Notably, the white-light light curve obtained on March 2014, which was taken under excellent conditions, attains a precision of $\sim 270$ ppm in 33-second exposures (giving a total cadence of $\sim60$ s, including the $30$ second readout time of the f/2 camera on \texttt{TURBO} mode). 
This is one of the best photometric precisions attained from the ground. 
Our other light curves are similarly precise, with the exception of the June 2015 data (which were taken under poor photometric conditions) and the April 12th 2017 data, which were taken with our ``blue" setup and, thus, on a portion of the spectrum in which (1) WASP-19 is not very bright and (2) the throughput of the instrument is not at its peak. 
We note that these observations are also more precise than both the {\em HST} ones for WASP-19, where \cite{Huitson:2013} reports a precision of 280 ppm with 293 second exposures with HST/WFC3, and the VLT/FORS2 ones, where \cite{FORSW19:2015} reports a precision of 697 ppm with 30 second exposures for WASP-19. 
Therefore, we believe that Magellan/IMACS will be an interesting option for very precise follow-up light curves of Transiting Exoplanet Survey Satellite \citep[TESS,][]{TESS:2014} targets.

Our precise white-light transit light curves allow us to refine the planetary parameters of the WASP-19b system. 
In Table~\ref{tab:system-params} we summarize the best-fit values for the transit light curve parameters for each of the observed nights. The retrieved parameters are all consistent within the errorbars between observations, which demonstrates the quality of both the precision and accuracy of our white-light transit light curves through different seasons and observing modes. Using the observed times of transit center, we 
obtain a revised ephemerides for WASP-19b (see details in Appendix 
\ref{sec:ephem}); in addition, we combine the orbital and physical 
parameters presented in Table~\ref{tab:system-params} and compare them 
to other works performing precise photometry for WASP-19b in Table~\ref{tab:final-system-params}. As 
can be seen, the agreement with the work of \cite{Mancini:2013} and 
\cite{Tregloan-Reed:2013} is, in general, excellent. We combine the precise planet-to-star radius ratios obtained in \cite{Mancini:2013} and in 
\cite{Tregloan-Reed:2013} with ours, obtaining a revised value for this 
parameter of $R_p/R_* = 0.14257 \pm 0.00027$. Using the stellar radius derived in \cite{Mancini:2013} of $R_s = 1.018 \pm 0.021R_\odot$ (where, for simplicity, we have added the systematic and random uncertainties) for WASP-19, we obtain a revised planetary radius of $R_p = 1.413\pm 0.029 R_J$. 
Using the mass for WASP-19b also derived in \cite{Mancini:2013}, we obtain a planetary gravity of $1414\pm 95$ cm/s$^2$. The importance of these revised parameters will be made evident in Section \ref{sec:discussion}.

\begin{table*}
    \centering
    \caption{Retrieved transit parameters obtained for each of our white-light transits; $s_1$ and $s_2$ stand for the first and second limb-darkening coefficient of the square-root law.}    
    \label{tab:system-params}
    \begin{tabular}{lcccccc} 
        \hline
        \hline
Date of observation & $t_0$ (BJD UTC) & $R_p/R_*$ & $a/R_*$ & $i$ (degs) & $s_1$ & $s_2$\\
        \hline
14/03/22 & $2456739.547178 \pm 0.000046$ & $0.14458 \pm 0.00041$ & $3.516 \pm 0.020$ & $78.98 \pm 0.15$ & $0.49 \pm 0.04$ & $0.05 \pm 0.05$ \\
14/04/29 & $2456776.622511 \pm 0.000066$ & $0.14371 \pm 0.00094$ & $3.480 \pm 0.046$ & $79.03 \pm 0.37$ & $0.09 \pm 0.15$ & $0.87 \pm 0.16$ \\
15/06/03 & $2457176.563965 \pm 0.000094$ & $0.14227 \pm 0.00105$ & $3.541 \pm 0.048$ & $79.16 \pm 0.38$ & $0.34 \pm 0.23$ & $0.31 \pm 0.40$ \\
17/02/11 & $2457796.591441 \pm 0.000061$ & $0.14225 \pm 0.00080$ & $3.621 \pm 0.034$ & $79.65 \pm 0.26$ & $0.49 \pm 0.07$ & $0.07 \pm 0.10$ \\
17/04/04 & $2457848.654791 \pm 0.000068$ & $0.14315 \pm 0.00074$ & $3.591 \pm 0.033$ & $79.72 \pm 0.26$ & $0.36 \pm 0.20$ & $0.27 \pm 0.34$ \\
17/04/12 & $2457856.543042 \pm 0.000111$ & $0.13803 \pm 0.00240$ & $3.694 \pm 0.074$ & $80.93 \pm 0.65$ & $0.80 \pm 0.08$ & $0.17 \pm 0.08$ \\
        \hline
        \hline
    \end{tabular}
\end{table*}

\begin{table*}
    \centering
    \caption{Final combined parameters for the orbit of WASP-19b from different references and this work. If the parameters are not directly in the references, they have been derived from the parameters fitted in the different works. $t_0$ 
    is the transit time at the first epoch, which is taken to be that of \protect\citet{Hebb:2010}.}    
    \label{tab:final-system-params}
    \begin{tabular}{lccccc} 
        \hline
        \hline
Reference  & $R_p/R_*$ & $a/R_*$ & $i$ (degs) & $t_0$ (BJD TDB) & Period (BJD TDB)\\  
\hline
\cite{Hebb:2010}  & $0.1424 \pm 0.0014$ & $3.87\pm 0.39$ & $80.80\pm 0.80$ & 2454775.33720(20) & 0.78883990(80) \\
\cite{Tregloan-Reed:2013} & $0.1428\pm 0.0006$ & $3.462 \pm 0.027$ & $78.94 \pm 0.23$ & 2454775.337548(18) & 0.78883942(33) \\
\cite{Mancini:2013}  & $0.14259\pm 0.00023$ & $3.4522\pm 0.0078$ & $78.76\pm 0.13$ & 2454775.33745(35) & 0.7888396(10) \\
\hline
\textbf{This work}  & $0.14233\pm 0.00050$ & $3.550 \pm 0.014$ & $79.29 \pm 0.10$ & 2454775.337777(42) & 0.788839316(17) \\
\hline 
        \hline
    \end{tabular}
\end{table*}

\subsection{Wavelength-dependent light curve analysis}
\label{sec:transpec}


\begin{figure}
   \includegraphics[height=1.5\columnwidth]{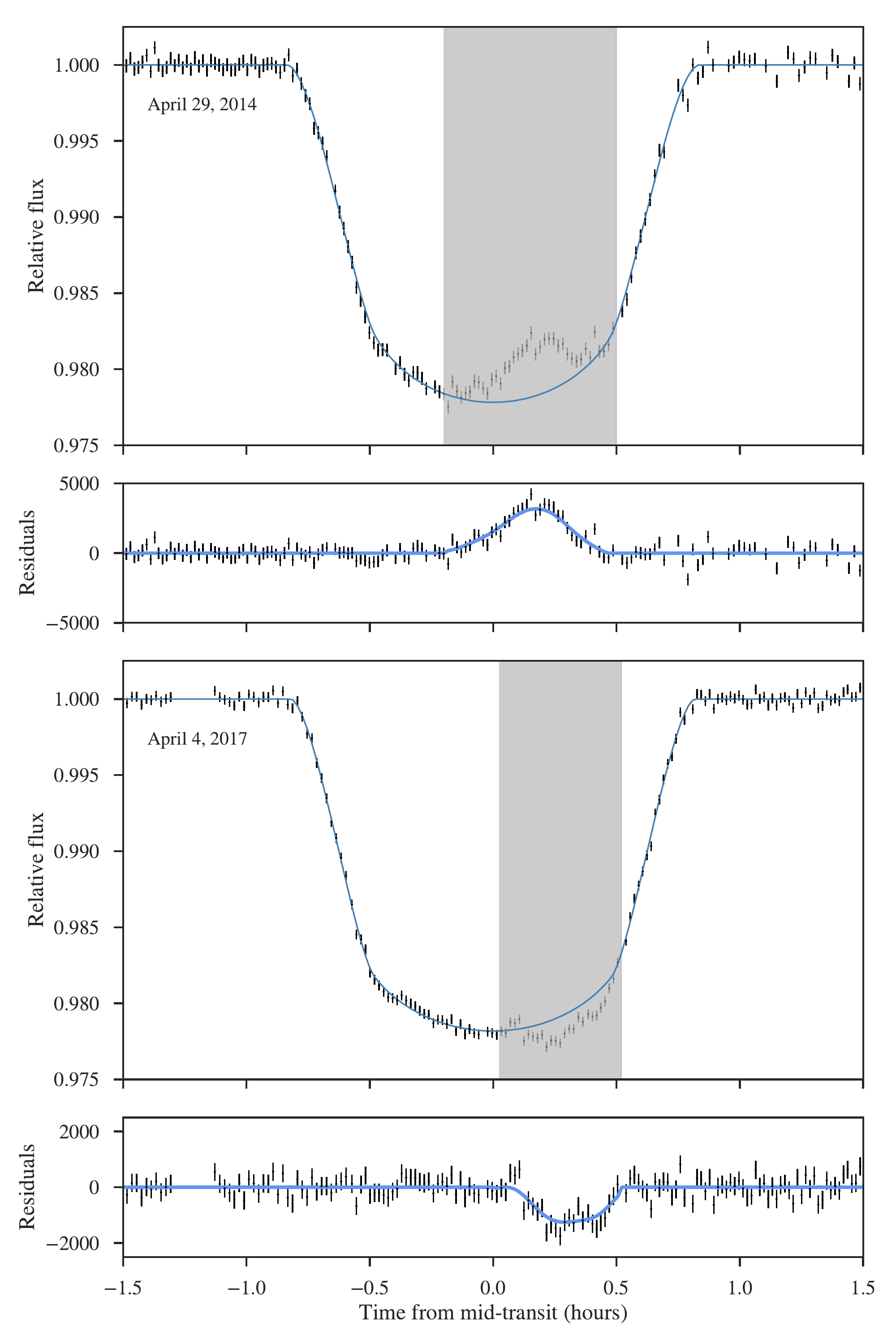}
    \caption{Spot shapes for the spot event on April 2014 (upper panels) and April 2017 (lower panels). The residuals (small panels below each transit light curve, in ppm) have been smoothed to extract the spot shape (blue curve in the residual panels), which was used to fit the wavelength-dependent transit light curves.}
    \label{fig:spots}
\end{figure}

For the analysis of the wavelength-dependent light curves, we followed a similar analysis as the one outlined 
for the white-light light curve analysis. However, we fix here the value of $P$, $a/R_*$, $i$, 
and $t_0$ to the values found in our white-light light curve analysis, so the only free 
parameters that define our wavelength-dependent transit light curves are the planet-to-star 
radius ratio, $R_p/R_*(\lambda)$, and the limb-darkening coefficient of the linear law used to account for this effect. For the systematics, we use the same model as in for the white-light analysis. 

For the light curves that show spot-crossing events, however, we apply a special treatment. 
Although proper analysis of the spot shapes, contrasts, and positions will be performed in the next section, for the measurement of the transmission spectrum here we subtract the predicted model of the white-light light curves from the observed transit containing the spot in order to obtain an estimate of the shape of the spot feature on our transit light curves. 
We use this estimate to correct the wavelength-dependent light curves in order to analyze all the light curves from different nights in a uniform way. 
To obtain this estimate from the residuals of our best-fit white-light light curve model, we apply a median filter with a 5-point window on the residuals, which we then smooth with a Gaussian filter with a 3-point 
standard deviation in order to get a smooth version of the spot shape. The result of this procedure is shown 
in Figure~\ref{fig:spots}. Using this spot shape, which we denote by $S(t)$, we model the transit at each 
wavelength as $T(t|\mathbf{\theta_T}) - AS(t)$, where $A$ is a term that modulates the amplitude of 
the spot event at each wavelength.
\begin{figure*}
   \includegraphics[height=1.7\columnwidth]{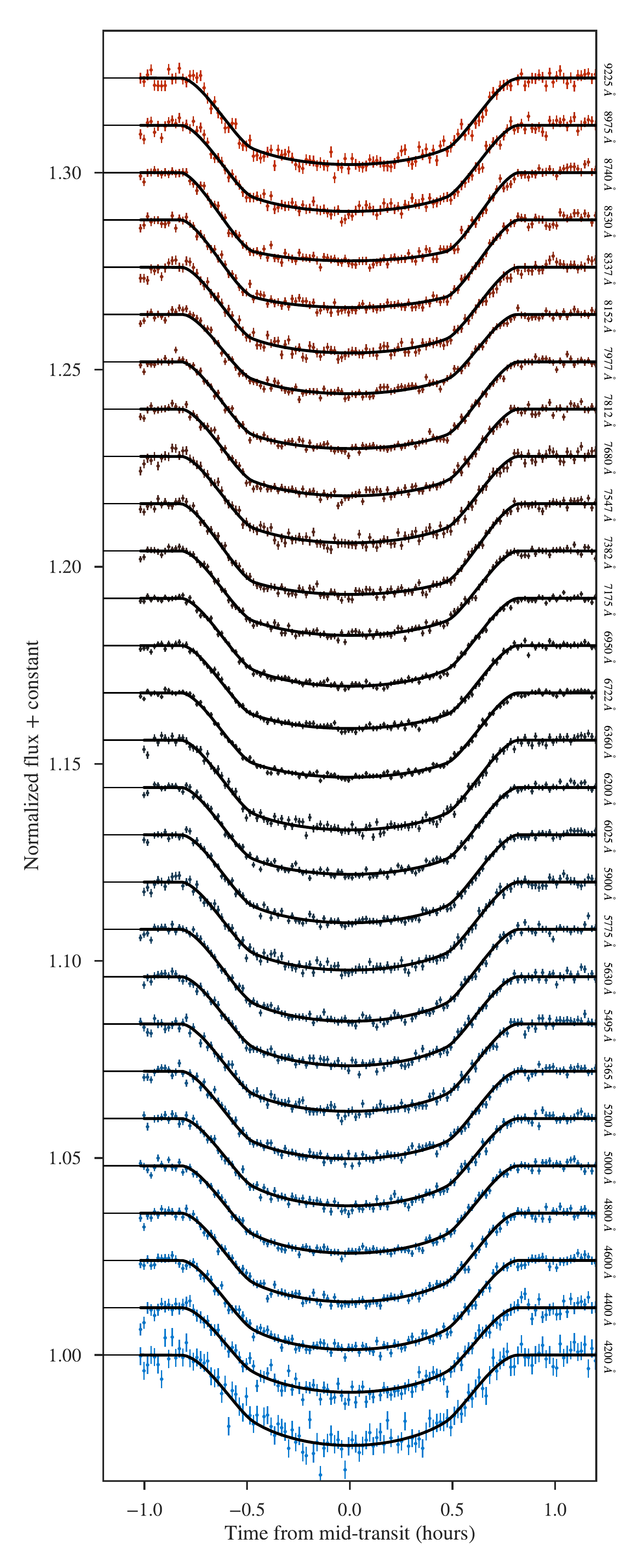}
   \includegraphics[height=1.7\columnwidth]{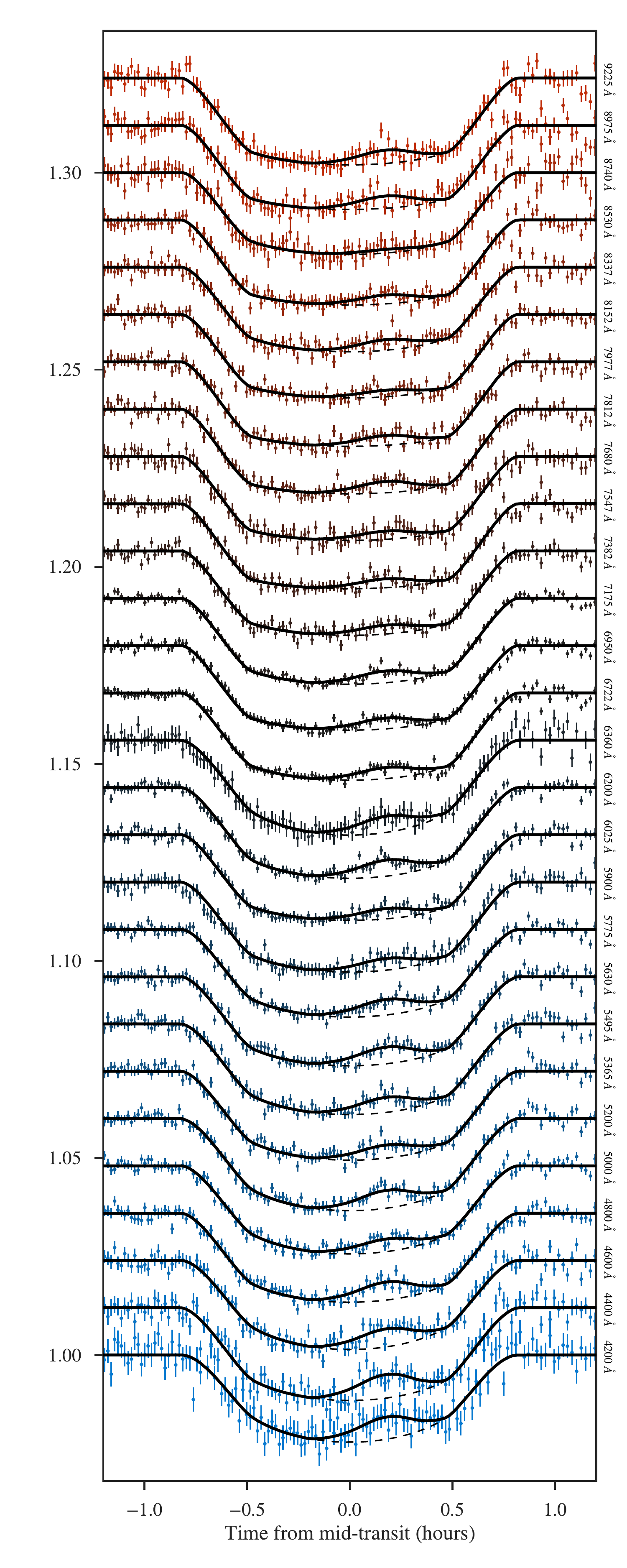}
   \includegraphics[height=1.7\columnwidth]{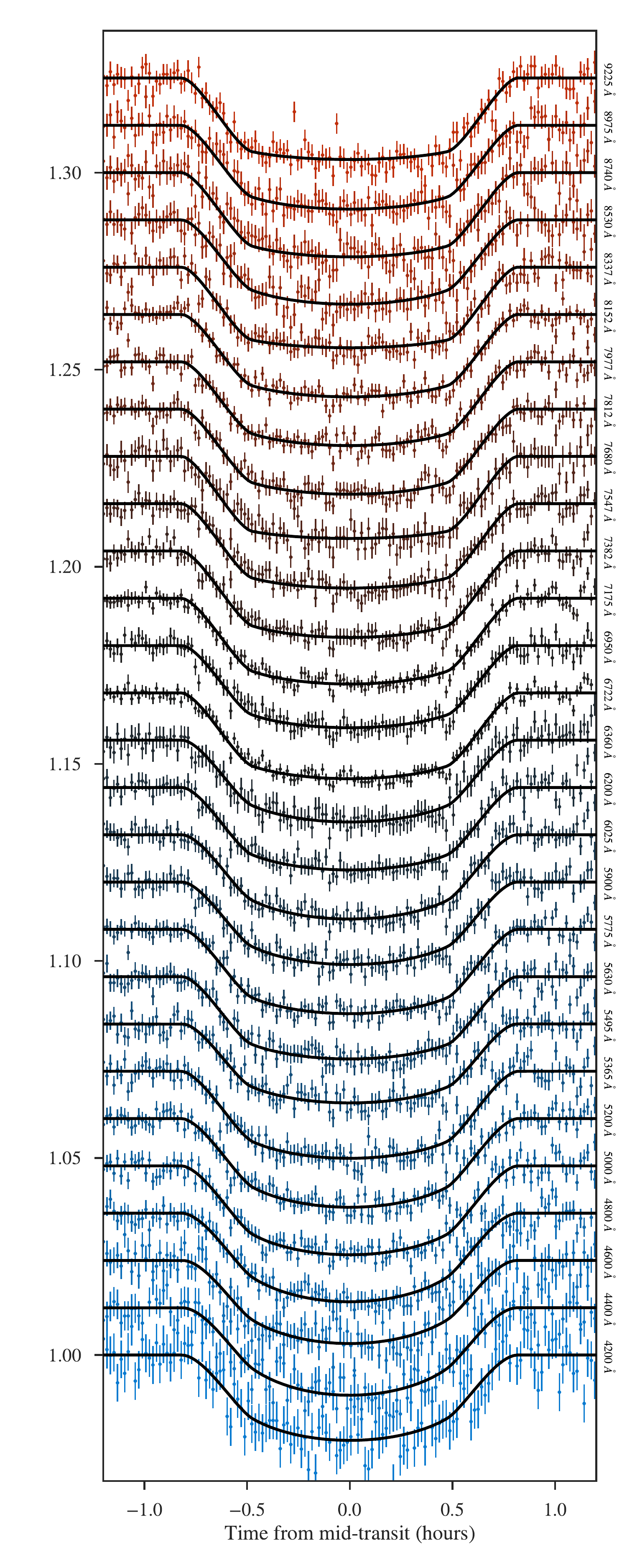}
    \caption{Transit light curves after our systematics removal for the 2014 and 2015 season (points with errorbars) along with the best-fit transit light curves (solid black lines): March, 2014 (left), April, 2014 (center) and June, 2015 (right). The mean wavelength of each bin is indicated to the right of each light curve. In the April 2014 panel (center), we show the transit model without the spot model with dashed lines.}
    \label{fig:wav-transits1}
\end{figure*}

\begin{figure*}
   \includegraphics[height=1.7\columnwidth]{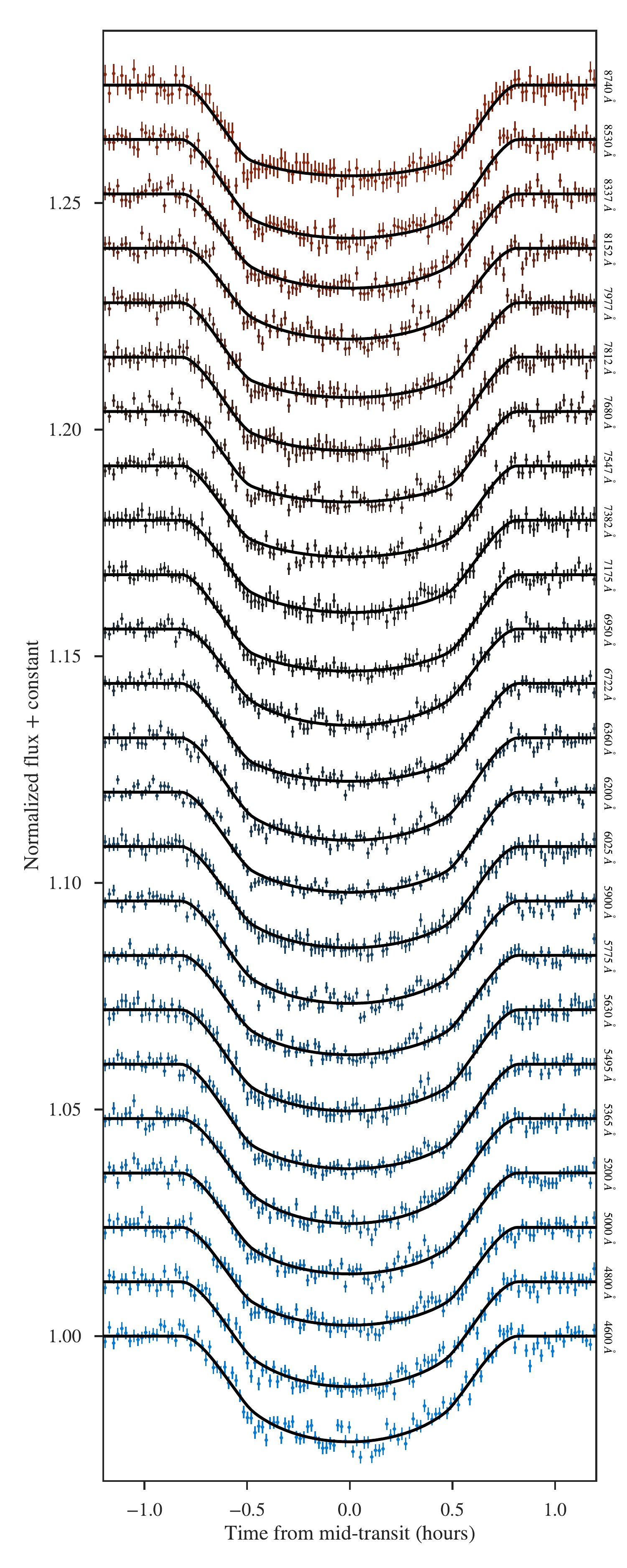}
   \includegraphics[height=1.7\columnwidth]{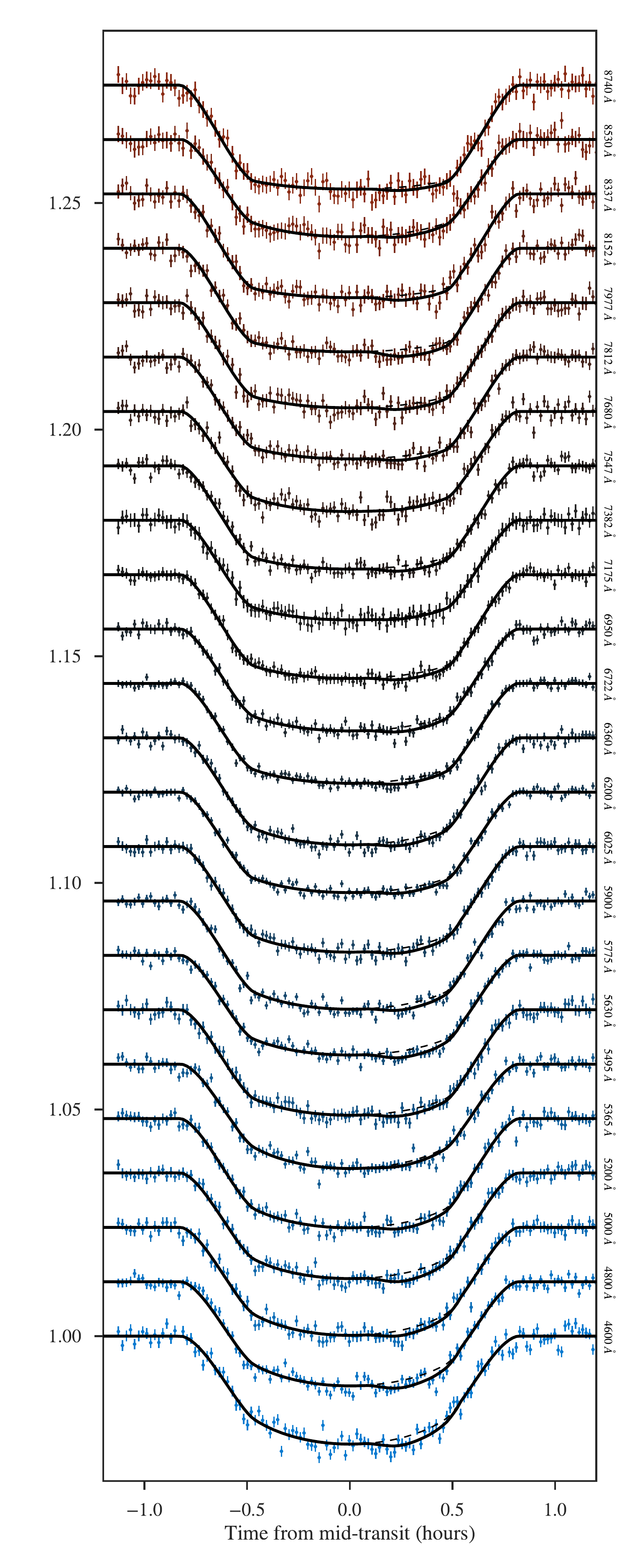}
   \includegraphics[height=1.7\columnwidth]{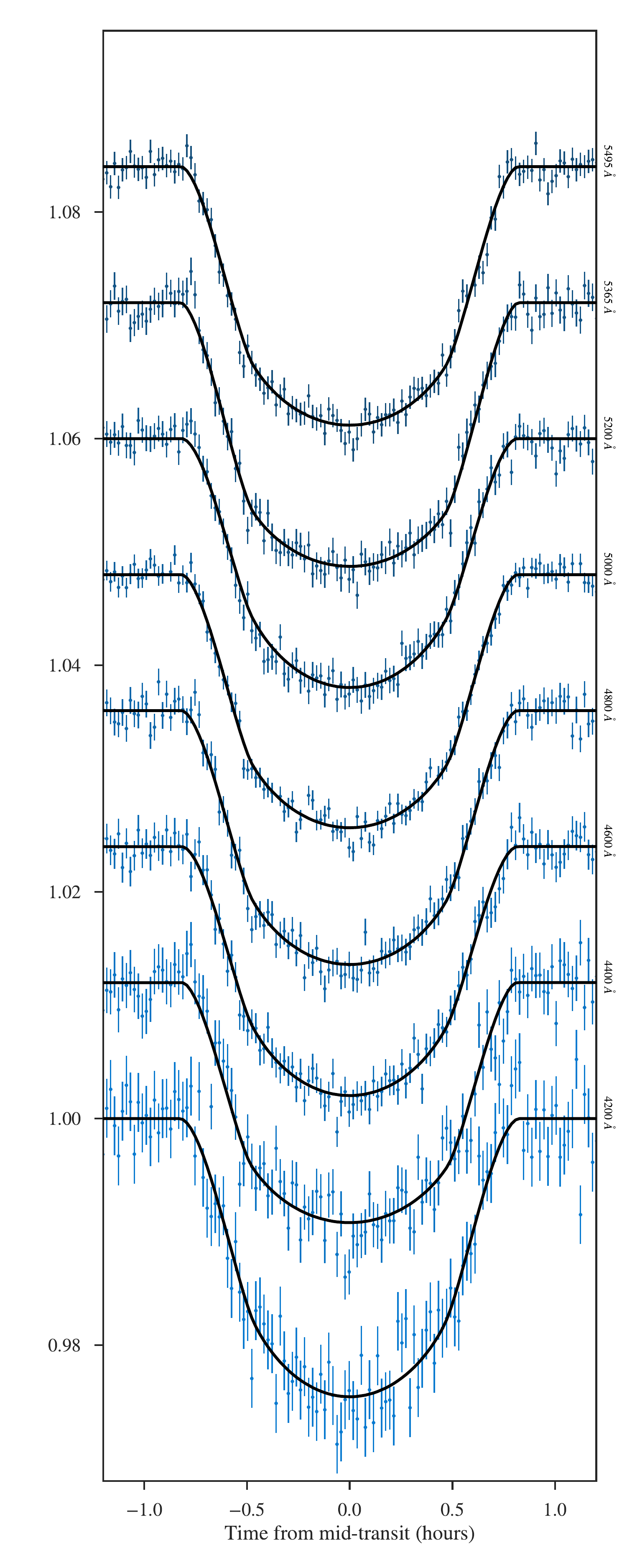}
    \caption{Transit light curves after our systematics removal for the 2017 season: February 2017 (left), April 4th, 2017 (center) and April 12th, 2017 (right). The mean wavelength of each bin is indicated to the right of each light curve. In the April 4th, 2017 panel (center), we show the transit model without the spot model with dashed lines. Note 
             the change in scale and wavelength range for the latter light curve, whose data was obtained with our ``blue" setup, and hence covers our bluest wavelength 
             bins.}
    \label{fig:wav-transits2}
\end{figure*}

Figures~\ref{fig:wav-transits1} and \ref{fig:wav-transits2} show the resulting light curves in the 
different wavelength bins analysed in this work.
The bin widths, detailed in Table~\ref{tab:transpec}, were selected to maximize the detectability of Na I and K I features as well as TiO/VO features (i.e., wide enough to have sufficient signal-to-noise in the light curves, but not too wide to lose spectral resolution). 
Our 2014 data span the wavelength range from 4100 to 9350 \AA\ in bins of variable size spanning from $\approx 150$ to $250$ \AA\ bins, which results in 28 bins across that wavelength range. 
Depending on the wavelength bin, the wavelength-dependent light curve jitter (i.e., $\sigma_w$ in our model) varies from 
around $\sim 600-1500$ ppm for the March data to around $\sim 900-2,000$ ppm for the April data. 
The 2015 data, taken with the same setup but during worse weather conditions, varies between $\sim 1,500-3,000$ ppm. 
For the 2017 season, the data in February and the 4th of April, taken with our ``red" setup, span a wavelength range from 4,500 to 9,350 \AA, in 26 wavelength bins. 
The wavelength-dependent light curves in our February and April 4th observations show jitter with values around $1000-1500$ and $800-1200$ ppm, respectively. 
Finally, the 2017 data taken on April 12th with our ``blue" setup, spans the wavelength range from 4,100 to 5,560 \AA, with jitter values close to $\approx 900$ ppm for the reddest wavelengths, reaching $\approx 2500$ ppm at the bluest wavelengths.

\begin{figure*}
\includegraphics[height=0.8\columnwidth]{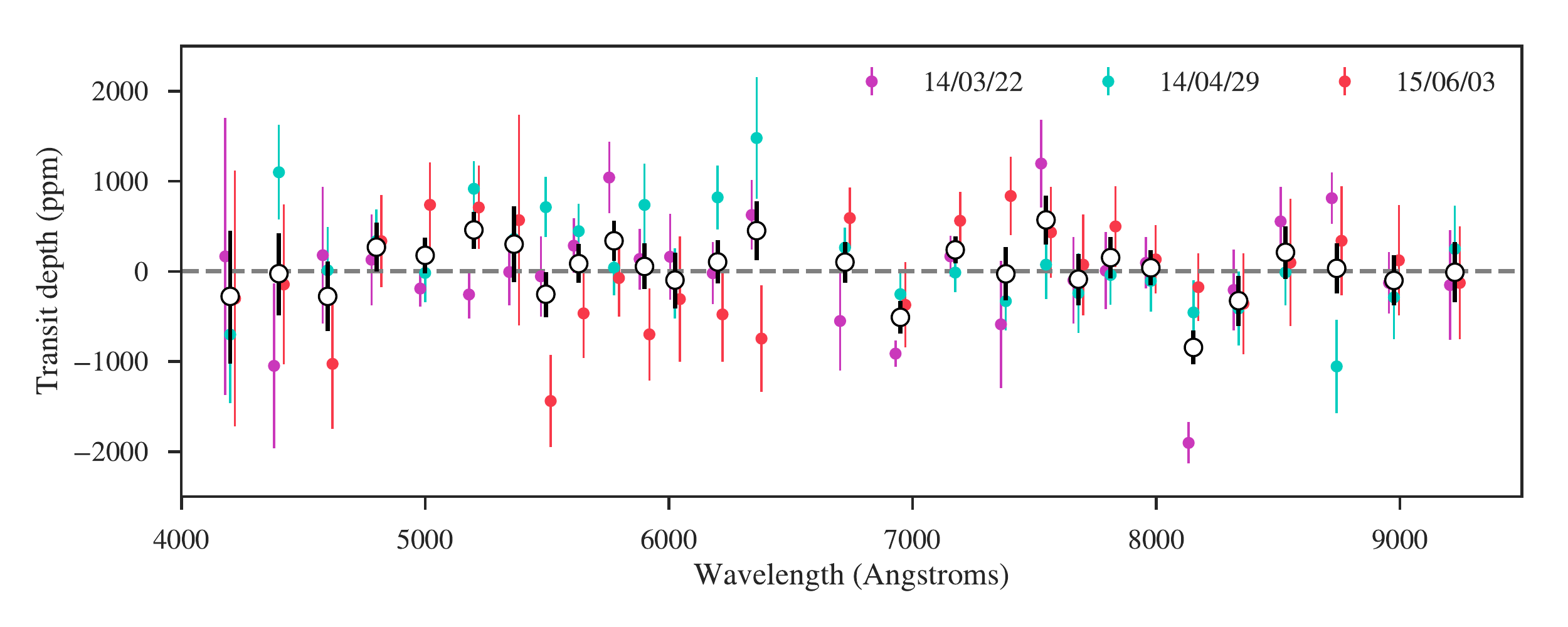}
\includegraphics[height=0.8\columnwidth]{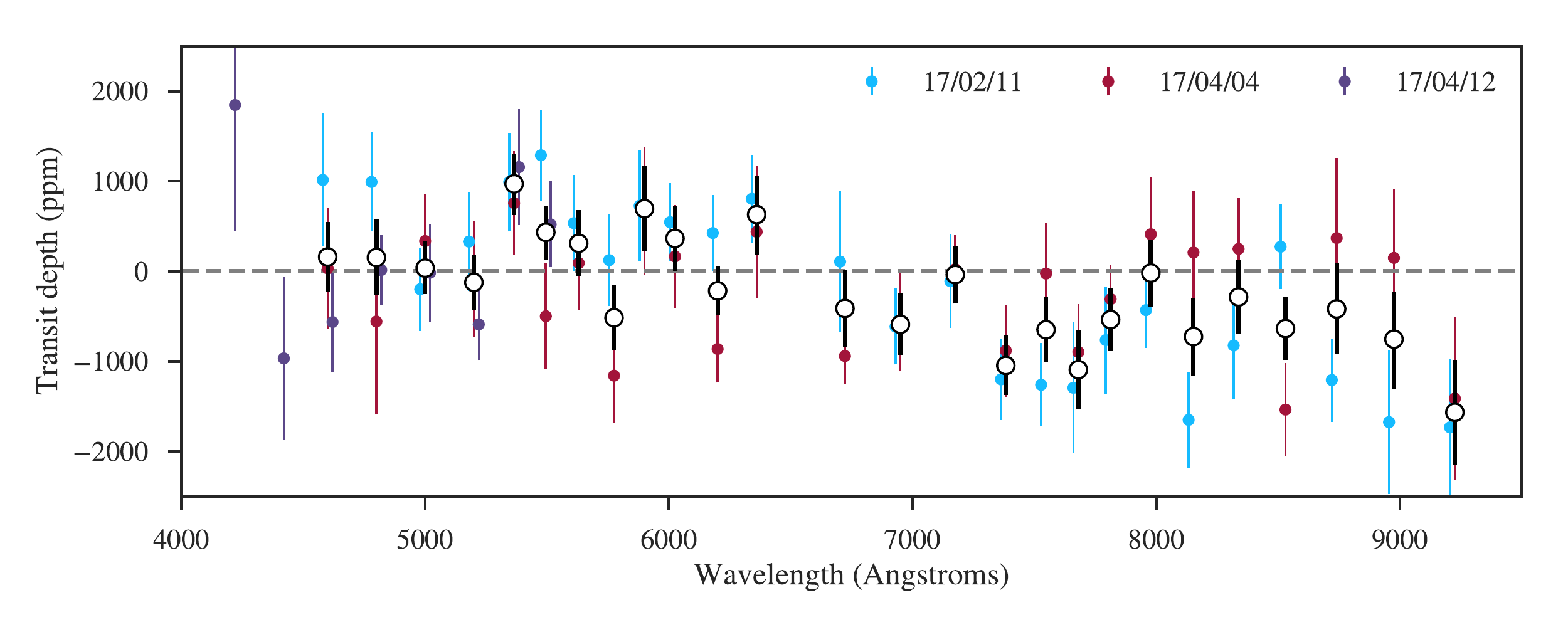}
    \caption{Combined transmission spectrum (white datapoints with errorbars) of WASP-19b obtained by combining the mean-subtracted transit depth of our 2014-2015 season (top, where no BF was used) and of the 2017 season (bottom, where a BF was used). Note the strong deviation of the 17/02/11 dataset from the behavior of all the other datasets, hinting to a strong decreasing slope as a function of wavelength.}
    \label{fig:transpec-epochs}
\end{figure*}

\begin{table*}
    \centering
    \caption{Transit depths in parts-per-million as a function of wavelength for the six 
    datasets presented in this work, along with the combined, mean-subtracted transmission 
    spectrum (last column, in bold) obtained by combining the mean-subtracted transmission 
    spectra of all nights but the 17/02/11 one (see text).}    
    \label{tab:transpec}
    \begin{tabular}{cccccccc} 
        \hline
        \hline
Wavelength (\AA)  & 14/03/22 & 14/04/29 & 15/06/03 & 17/02/11 & 17/04/04 & 17/04/12 & \textbf{Combined}\\
        \hline
4100-4300  & $20973 \pm 1535$ &  $20095 \pm 759$ &  $19873 \pm 1418$ &  --- &  --- &  $22417 \pm 1394$ &  $\mathbf{269^{+652}_{-666}}$ \\
4300-4500  & $19758 \pm 917$ &  $21896 \pm 524$ &  $20029 \pm 886$ &  --- &  --- &  $19606 \pm 909$ &  $\mathbf{-261^{+403}_{-414}}$ \\
4500-4700  & $20986 \pm 760$ &  $20808 \pm 481$ &  $19148 \pm 724$ &  $21175 \pm 740$ &  $22019 \pm 676$ &  $20009 \pm 550$ &  $\mathbf{-270^{+286}_{-290}}$ \\
4700-4900  & $20936 \pm 503$ &  $21134 \pm 348$ &  $20510 \pm 512$ &  $21153 \pm 550$ &  $21430 \pm 1032$ &  $20586 \pm 387$ &  $\mathbf{53^{+276}_{-270}}$ \\
4900-5100  & $20616 \pm 198$ &  $20779 \pm 321$ &  $20912 \pm 472$ &  $19962 \pm 460$ &  $22322 \pm 525$ &  $20556 \pm 545$ &  $\mathbf{170^{+193}_{-196}}$ \\
5100-5300  & $20550 \pm 269$ &  $21713 \pm 305$ &  $20884 \pm 462$ &  $20491 \pm 547$ &  $21904 \pm 642$ &  $19983 \pm 395$ &  $\mathbf{137^{+196}_{-197}}$ \\
5300-5430  & $20800 \pm 370$ &  $21156 \pm 316$ &  $20742 \pm 1170$ &  $21152 \pm 547$ &  $22744 \pm 576$ &  $21728 \pm 644$ &  $\mathbf{570^{+299}_{-310}}$ \\
5430-5560  & $20751 \pm 445$ &  $21509 \pm 334$ &  $18736 \pm 513$ &  $21449 \pm 508$ &  $21488 \pm 588$ &  $21092 \pm 476$ &  $\mathbf{-149^{+207}_{-215}}$ \\
5560-5700  & $21092 \pm 302$ &  $21242 \pm 307$ &  $19708 \pm 493$ &  $20696 \pm 536$ &  $22077 \pm 514$ &  --- &  $\mathbf{88^{+207}_{-208}}$ \\
5700-5850  & $21848 \pm 394$ &  $20837 \pm 303$ &  $20100 \pm 424$ &  $20285 \pm 510$ &  $20828 \pm 525$ &  --- &  $\mathbf{-36^{+214}_{-210}}$ \\
5850-5950  & $20943 \pm 338$ &  $21534 \pm 459$ &  $19475 \pm 512$ &  $20890 \pm 610$ &  $22655 \pm 715$ &  --- &  $\mathbf{206^{+261}_{-259}}$ \\
5950-6100  & $20969 \pm 476$ &  $20661 \pm 390$ &  $19866 \pm 694$ &  $20707 \pm 435$ &  $22151 \pm 571$ &  --- &  $\mathbf{-27^{+262}_{-281}}$ \\
6100-6300  & $20786 \pm 344$ &  $21617 \pm 353$ &  $19697 \pm 524$ &  $20586 \pm 419$ &  $21124 \pm 374$ &  --- &  $\mathbf{-134^{+194}_{-199}}$ \\
6300-6420  & $21433 \pm 385$ &  $22276 \pm 672$ &  $19429 \pm 591$ &  $20966 \pm 489$ &  $22425 \pm 732$ &  --- &  $\mathbf{448^{+299}_{-307}}$ \\
6595-6850  & $20256 \pm 547$ &  $21059 \pm 222$ &  $20766 \pm 336$ &  $20269 \pm 785$ &  $21047 \pm 311$ &  --- &  $\mathbf{-160^{+187}_{-184}}$ \\
6850-7050  & $19894 \pm 145$ &  $20543 \pm 260$ &  $19802 \pm 471$ &  $19548 \pm 421$ &  $21433 \pm 555$ &  --- &  $\mathbf{-527^{+199}_{-196}}$ \\
7050-7300  & $20973 \pm 226$ &  $20783 \pm 217$ &  $20736 \pm 319$ &  $20053 \pm 520$ &  $22010 \pm 379$ &  --- &  $\mathbf{186^{+147}_{-147}}$ \\
7300-7465  & $20219 \pm 705$ &  $20466 \pm 323$ &  $21011 \pm 434$ &  $18962 \pm 448$ &  $21106 \pm 511$ &  --- &  $\mathbf{-242^{+263}_{-257}}$ \\
7465-7630  & $22003 \pm 487$ &  $20871 \pm 381$ &  $20608 \pm 502$ &  $18902 \pm 461$ &  $21961 \pm 568$ &  --- &  $\mathbf{417^{+240}_{-241}}$ \\
7630-7730  & $20707 \pm 479$ &  $20555 \pm 440$ &  $20249 \pm 560$ &  $18869 \pm 724$ &  $21090 \pm 529$ &  --- &  $\mathbf{-285^{+249}_{-252}}$ \\
7730-7895  & $20813 \pm 428$ &  $20757 \pm 332$ &  $20673 \pm 449$ &  $19398 \pm 593$ &  $21678 \pm 374$ &  --- &  $\mathbf{40^{+192}_{-197}}$ \\
7895-8060  & $20901 \pm 284$ &  $20690 \pm 344$ &  $20307 \pm 380$ &  $19731 \pm 422$ &  $22397 \pm 630$ &  --- &  $\mathbf{134^{+211}_{-212}}$ \\
8060-8245  & $18903 \pm 230$ &  $20340 \pm 358$ &  $19999 \pm 373$ &  $18512 \pm 535$ &  $22195 \pm 684$ &  --- &  $\mathbf{-583^{+220}_{-224}}$ \\
8245-8430  & $20600 \pm 446$ &  $20384 \pm 412$ &  $19816 \pm 560$ &  $19338 \pm 599$ &  $22235 \pm 567$ &  --- &  $\mathbf{-184_{+248}^{+247}}$ \\
8430-8630  & $21361 \pm 382$ &  $20784 \pm 362$ &  $20271 \pm 705$ &  $20435 \pm 470$ &  $20451 \pm 517$ &  --- &  $\mathbf{-226^{+257}_{-248}}$ \\
8630-8850  & $21619 \pm 285$ &  $19741 \pm 517$ &  $20514 \pm 604$ &  $18955 \pm 463$ &  $22356 \pm 889$ &  --- &  $\mathbf{113^{+309}_{-304}}$ \\
8850-9100  & $20678 \pm 342$ &  $20508 \pm 461$ &  $20297 \pm 612$ &  $18487 \pm 796$ &  $22135 \pm 766$ &  --- &  $\mathbf{-35^{+281}_{-293}}$ \\
9100-9350  & $20654 \pm 609$ &  $21045 \pm 480$ &  $20047 \pm 626$ &  $18429 \pm 753$ &  $20575 \pm 900$ &  --- &  $\mathbf{-360^{+327}_{-321}}$ \\
        \hline
    \end{tabular}
\end{table*}

As discussed in Section \ref{sec:phot-mon}, WASP-19 is an active star and as such, differences in both the mean and wavelength-dependent transit depths are expected due to unocculted stellar spots. Differences between the retrieved mean transit depths are indeed observed in our data, with typical differences on the order of $\sim 500$ ppm, except for the 17/04/04 dataset, where a difference of $\sim 1000$ ppm is observed. 
Because of this, we work in what follows with the mean-subtracted transit depths, although we provide a detailed analysis of each dataset in Section \ref{sec:discussion}. 
In Table \ref{tab:transpec} we present the transmission spectra from all our epochs.
Figure \ref{fig:transpec-epochs} illustrates these spectra, grouped by whether we used a BF (which was the case for the 2014-2015 season) or not (which was the case for the data obtained from the 2017 season). 
Overall, almost all the datasets are consistent with a flat line within the errorbars, which suggests that whether a BF is used does not have a large impact at the level of precision of our Magellan/IMACS observations. 

The only exception to the observed featureless spectra is the February 2017 dataset, which shows an apparent slope that decreases as a function of wavelength with an amplitude of about 3000 ppm. 
Interestingly, this dataset was taken during a period of larger photometric variation for WASP-19, according to our analysis in \ref{sec:phot-mon-2017}. 
Cold unocculted spots covering a significant fraction of the star could produce such a slope, which to first order, should make the star appear dimmer and redder.
However, depending on their distribution and sizes, spots on the star can emulate a variety of photometric signatures which do not have a 1-to-1 correspondence with the spotted area, which complicates determinations of the stellar contamination signal from photometric monitoring alone \citep{TLSE:2018}.
Alternatively, the crossing of bright areas by the exoplanet could also lead to the same effect \citep{Oshagh:2014}. 
We discuss the impact of stellar heterogeneities on our transmission spectrum in detail in Section \ref{analysis:heter} and provide evidence that this indeed can explain the observed variation in our February 2017 dataset in Section \ref{sec:discussion}.

\begin{figure*}
   \includegraphics[height=0.8\columnwidth]{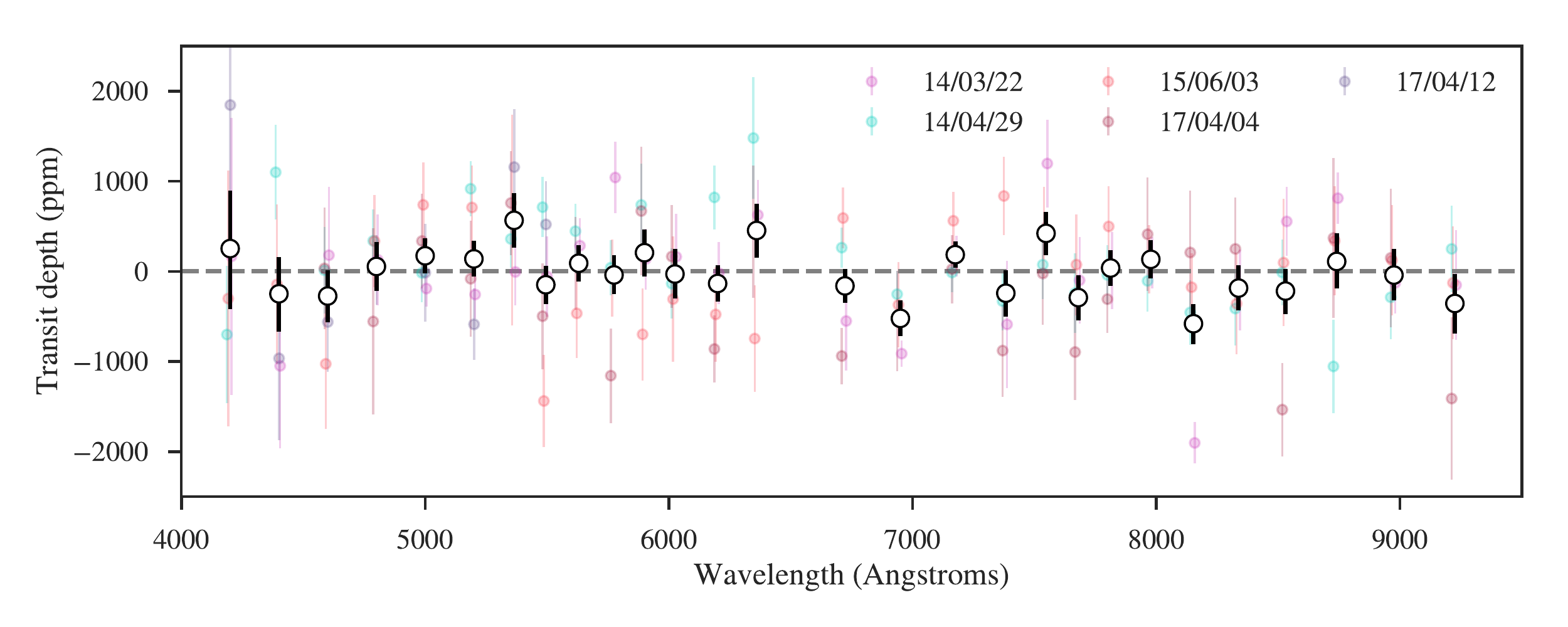}
    \caption{Final combined transmission spectrum (white datapoints with errorbars) of WASP-19b obtained with our Magellan/IMACS observations by combining the mean-subtracted transit depth of all of our epochs (colored datapoints with errorbars) except for the 17/02/11 dataset, which showed a strong decreasing slope with wavelength, which we here attribute to stellar activity. The mean transit depth of the combined transmission spectrum is $-27 \pm 44$ ppm.}
    \label{fig:transpec}
\end{figure*}

In Figure \ref{fig:transpec} we combine all of our measurements except for the February 2017 data (due to the observed slope) in order to generate our combined transmission spectrum. 
The resulting spectrum is consistent with a flat line, with a mean transit depth of $-27 \pm 44$ ppm. 
If we assume this (null) hypothesis to be true, then the probability that we observed these data or more extreme values (i.e., the p-value) using our 28 datapoints (which imply 27 degrees of freedom, given our estimated mean) is $0.15$, with which we cannot reject the hypothesis that the data actually comes from a simple flat line. 
Of course, this simple null-hypothesis testing scheme ignores the possibility of other possible hypotheses that could explain the data. 
We discuss those in the next section. 

\subsection{Spot-crossing events}
\label{sec:spot-crossing}


\begin{figure*}
   \includegraphics[height=1.3\columnwidth]{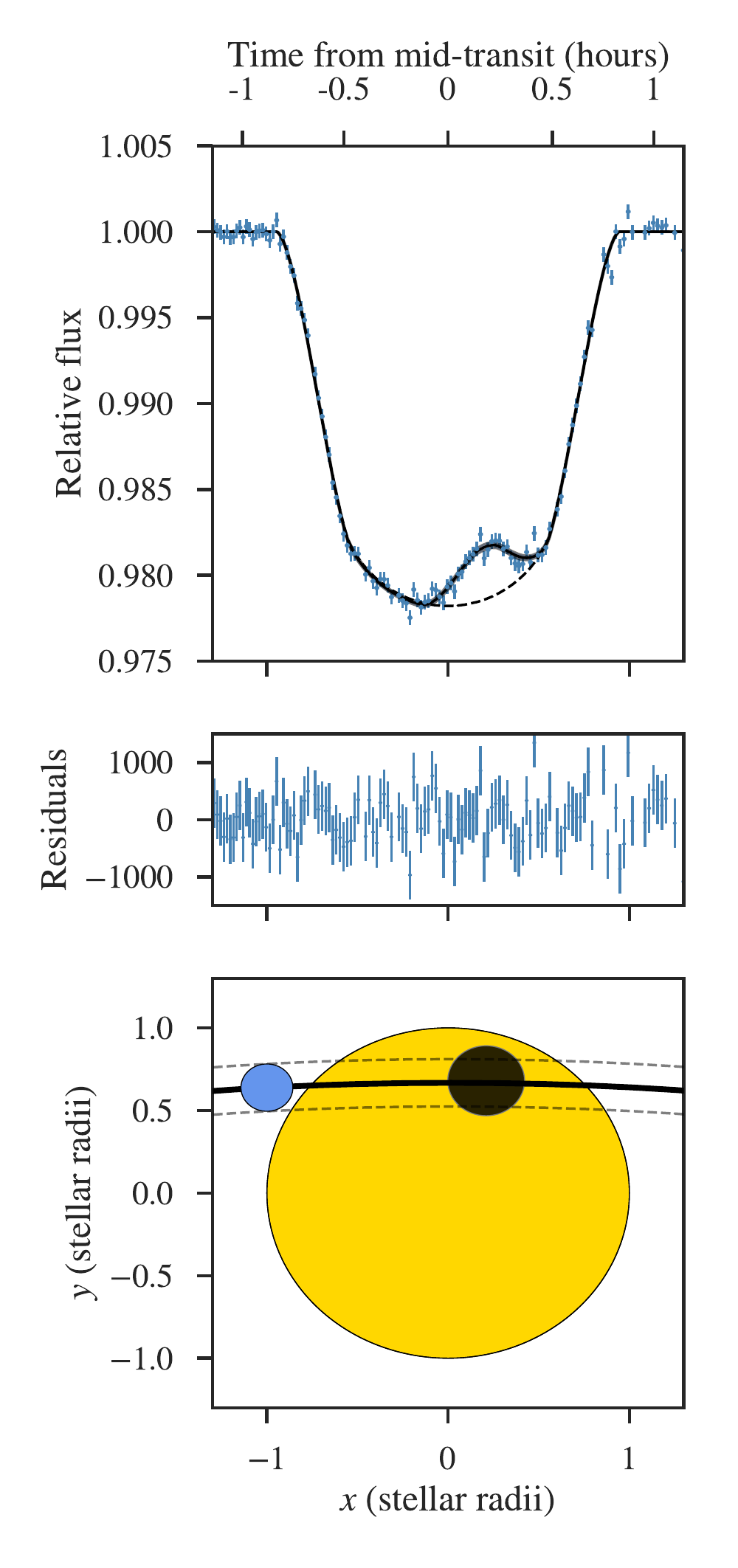}
   \includegraphics[height=1.3\columnwidth]{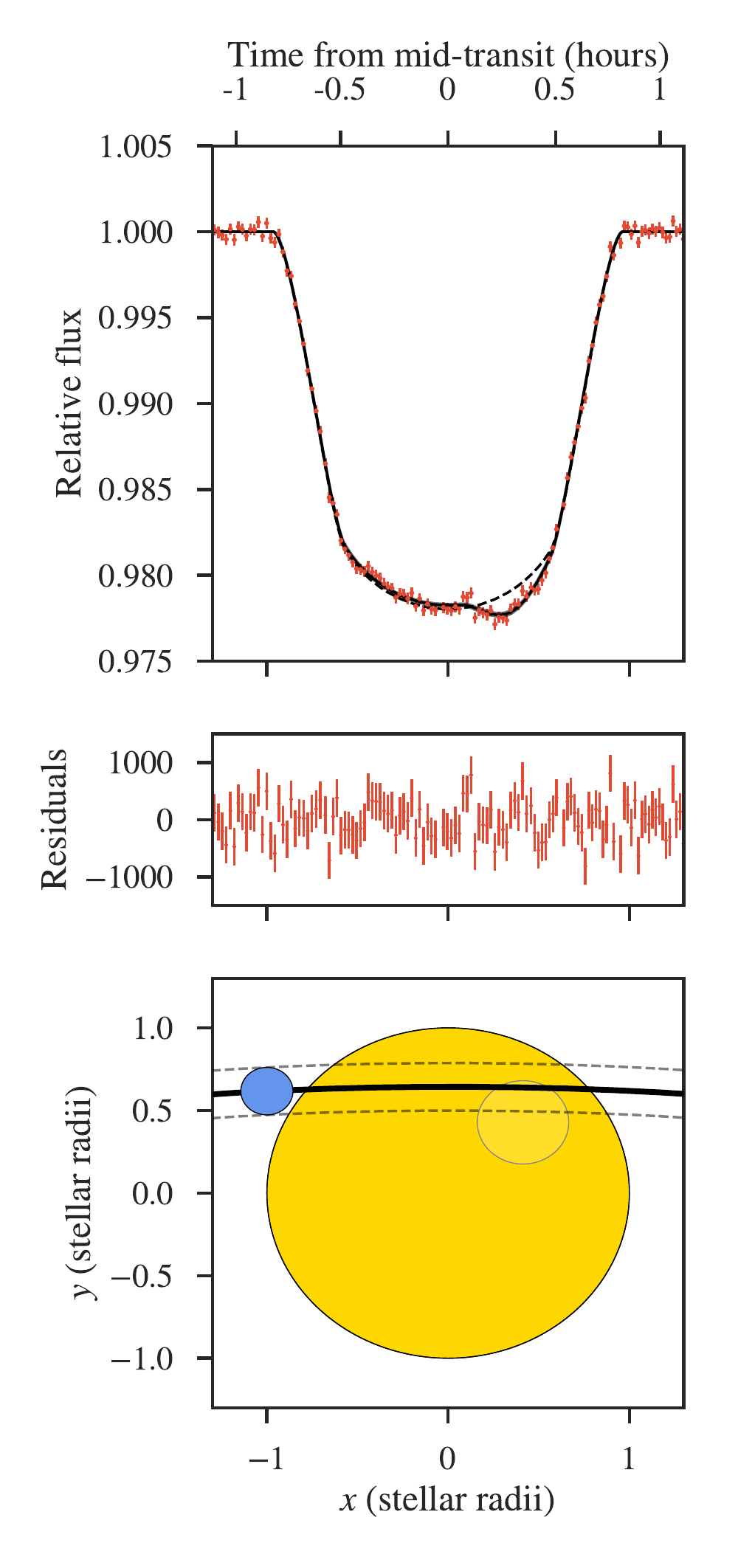}
   \includegraphics[height=1.3\columnwidth]{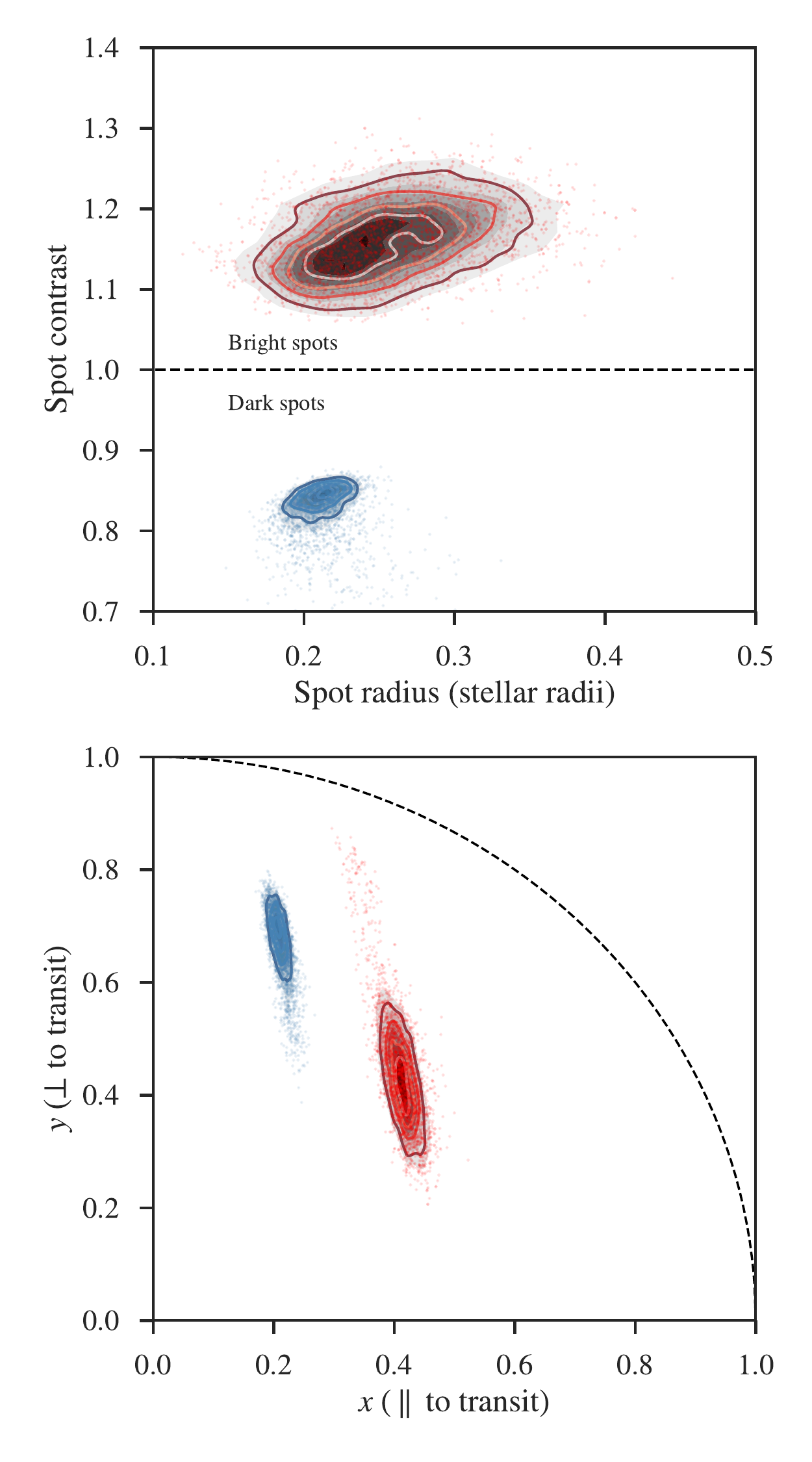}
\caption{Spot-crossing event modelling for the April 2014 (left) and April 2017 (center) datasets. The top panels show the light curve distortions (blue for the 2014 event, red for the 2017 event) along with the best-fit model in black. The middle panels show the residuals, while in the bottom panel we present a physical representation of the spots (black and grey circles, respectively, to scale in the plot both in size and contrast) on the star (big yellow circle), along with the transit motion (blue circle representing the planet, its motion represented by the black line and with black dashed lines representing the transit chord). The right panels show the samples from the posterior distribution of the parameters (radius and contrast in the top panel, position on the stellar surface in the bottom panel, where in this latter panel we also indicate the stellar surface with dashed lines) of the spots for both datasets, where the posterior for each dataset is color coded with the same colors as the left and center panels.}
    \label{fig:spot-wl}
\end{figure*}

The spot-crossing events mentioned in the previous section for the April 2014 and 2017 data were analyzed using \texttt{spotrod} \citep{spotrod}, which allows us to 
fit for multiple spots on the stellar surface simultaneously with the transit parameters. Because the posterior 
sampling is a complicated (possibly multi-modal) function, we here use MultiNest \citep{MultiNest} via the PyMultiNest package \citep{PyMultiNest} to both (1) sample points from the posteriors and (2) compute posterior bayesian evidences, $Z = \mathbb{P}(D|H)$, i.e., the probability of the data $D$ given hypothesis $H$. This latter property of nested sampling algorithms in turn 
allows us to study how complex our models have to be in order to explain the observed light curve distortions (e.g., $n$ spots on the surface) via the posterior odds, $\mathbb{P}(H_n|D)/\mathbb{P}(H_k|D)$, where $\mathbb{P}(H_n|D) = \mathbb{P}(D|H_n)\mathbb{P}(H_n)$, with $\mathbb{P}(H_n)$ being the prior probability on the hypothesis $H_n$. 
Note that in the case in which the prior probability of the models is the same, 
then the posterior odds are simply the ratio of the evidences, 
$\mathbb{P}(H_n|D)/\mathbb{P}(H_k|D) = Z_n/Z_k$. 
In this work, we usually 
work in terms of the differences of the log-evidences, $\Delta \ln Z$, which are the logarithm of the odds, assuming equal prior probabilities for the hypotheses. 
A good review in terms of how the posterior odds translate to frequentist 
significance hypothesis testing can be found in \cite{Trotta:2008}; we note from that work that absolute values of log-odds below 1 are usually considered inconclusive, whereas absolute log-odds of 2.5 can be interpreted as moderate evidence and absolute log-odds higher than 5 can be interpreted usually as highly significant. We caution, however, that this comparison with frequentist methods is usually useful for understanding but not very meaningful in practice in the 
sense that frequentist hypothesis testing has only one null hypothesis, whereas proper bayesian model comparison considers a range of possible hypotheses at play. The engine for modelling transiting exoplanet light curves using \texttt{spotrod} with PyMultinest, \texttt{spotnest}, is available at GitHub\footnote{\url{http://github.com/nespinoza/spotnest}}. 

A detailed explanation of the analysis of the spot-crossing events can be found in Appendix \ref{sec:spot-analysis}; we here discuss the main results of that analysis.
For the April 2014 dataset, using the white-light light curve we found that the best model 
for the spot-crossing event is one with a single spot located at $(x,y)=(0.21^{+0.01}_{-0.01},0.68^{+0.04}_{-0.06})$, where $x$ goes in the direction almost perpendicular to the planet's motion and the center of the star is at $(0,0)$.
The spot has a contrast of $F_\textnormal{spot}/F_\textnormal{star} = 0.84^{+0.01}_{-0.03}$, (i.e., a dark spot) and a considerably large radius of $R_\textnormal{spot} = 0.21^{+0.01}_{-0.01}$ stellar radii. 
For the April 2017 dataset, the one-spot model is also favored; the retrieved position of the spot is 
identified with high precision to be at $(x,y)=(0.41^{+0.02}_{-0.02},0.43^{+0.08}_{-0.08})$ on the stellar surface, with a size of $0.25^{+0.05}_{-0.04}$ stellar radii. 
The most interesting feature of the April 2017 feature, however, is its contrast: the retrieved spot contrast is $F_\textnormal{spot}/F_\textnormal{star} = 1.16^{+0.05}_{-0.04}$, which implies the feature is actually a \textit{bright} spot. 
This makes the April 2017 event, thus, one of the first unambiguously detected bright spot features, i.e. a facular or plage region, on an exoplanet host star \citep[along with that of][ on WASP-52]{firstfac:2016} and the first one on WASP-19. 
A physical depiction of the star-spot-planet system is shown in Figure \ref{fig:spot-wl}, along with the posterior distributions of the spot contrasts, sizes and positions retrieved from our light curve analysis.

\begin{figure}
   \includegraphics[height=0.7\columnwidth]{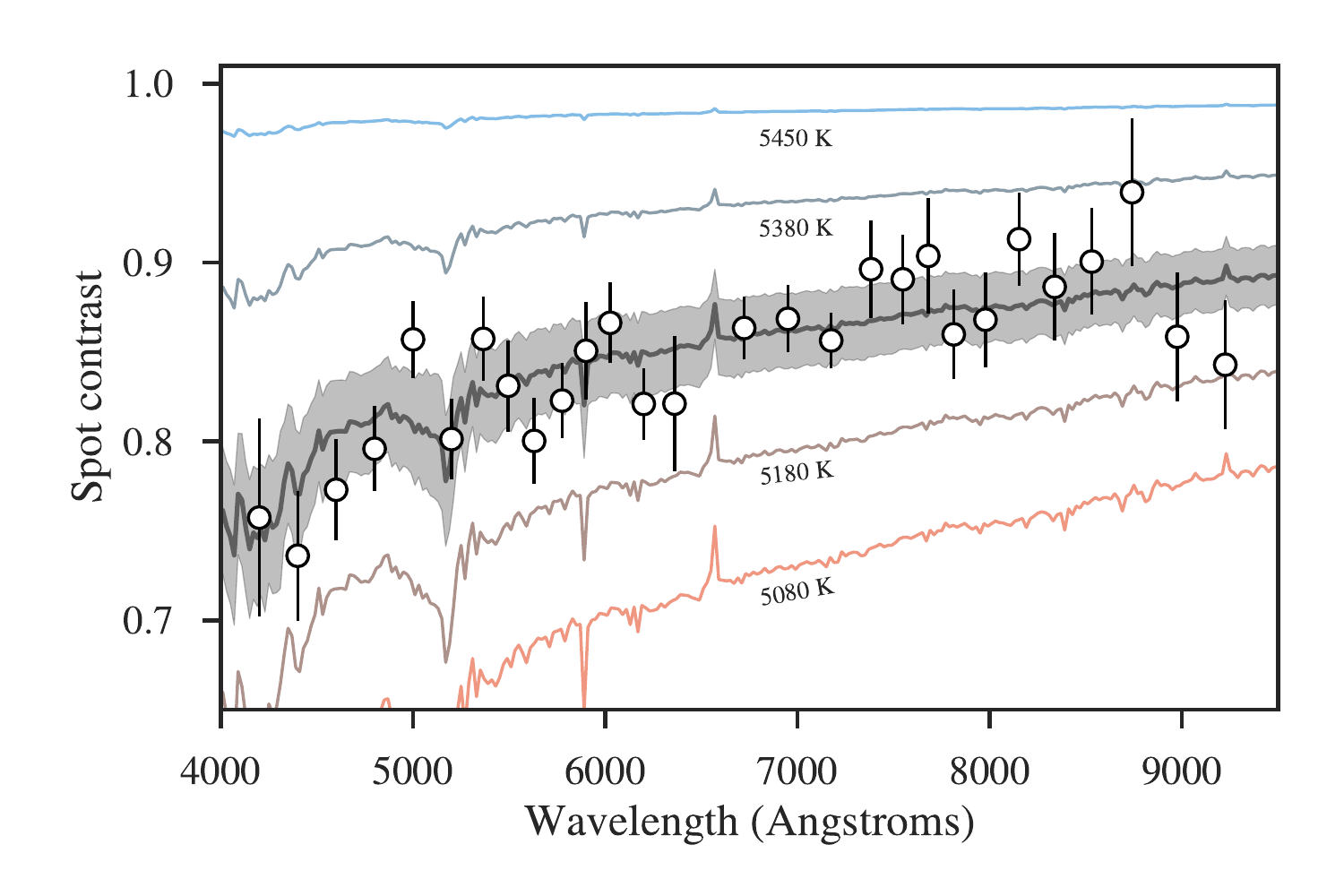}
   \includegraphics[height=0.7\columnwidth]{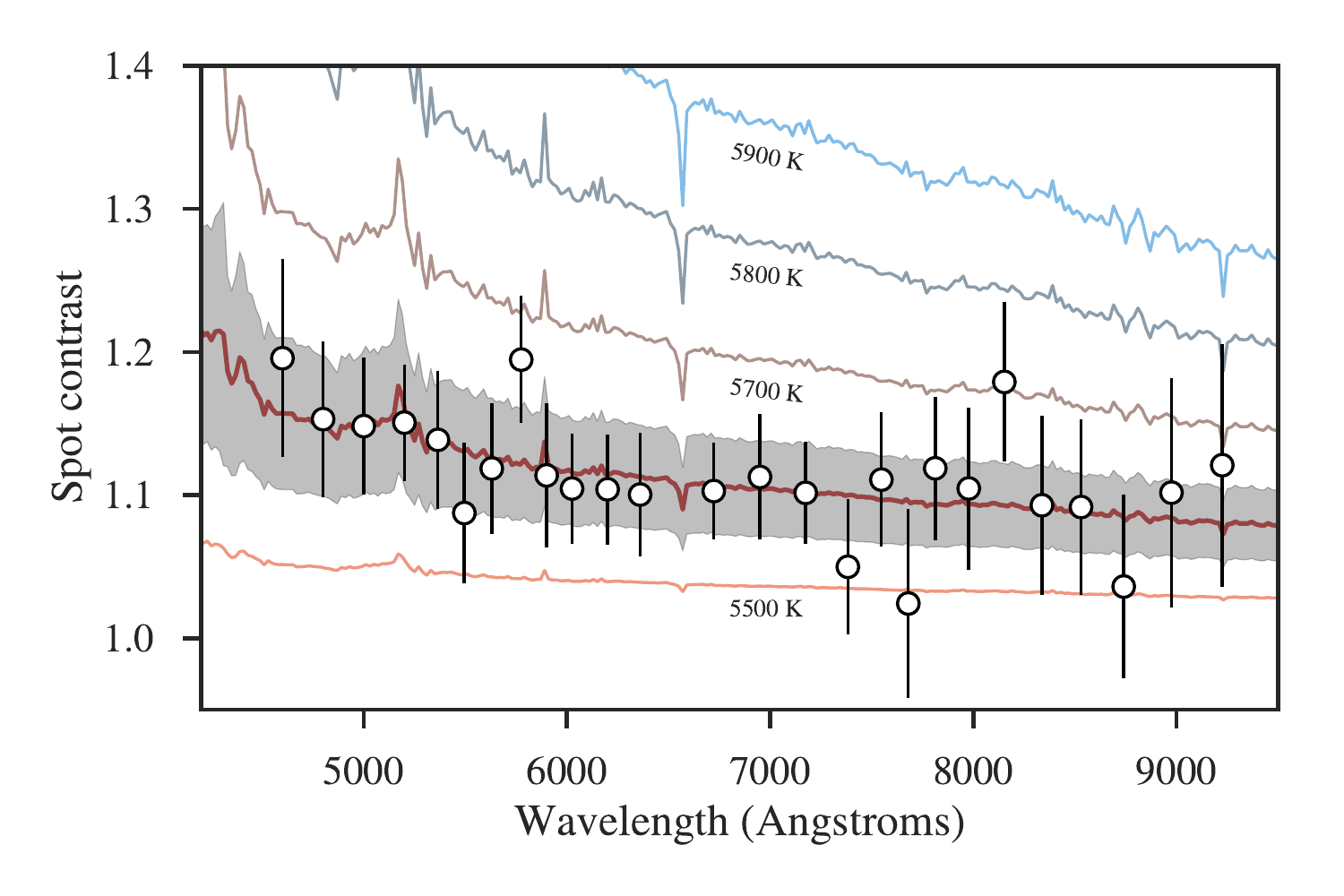}
\caption{Contrasts as a function of wavelength obtained by our wavelength-dependent light curve modelling of the spot features, along with our best-fit models (lines with grey bands, which represent the 5-sigma credibility band of our best-models) for the spot temperatures for our April 2017 (lower panel) and 2014 (upper panel) datasets. For the bright spot (April 2017 dataset), the best-fit temperature is $T_s = 5588 \pm 92$ K (or a $\Delta T = 137 \pm 10$ K with respect to the star); for the cool spot (April 2014 dataset), the best-fit temperature is $T_s = 5278 \pm 81$ K (or a $\Delta T = 192 \pm 10$ K colder than the star). Colored lines represent models for 
different spot temperatures for comparison.}
    \label{fig:contrast}
\end{figure}

Using the derived shapes and positions of the spots from the white-light light curves, we analysed the systematics-removed, wavelength-dependent light curves to study the spot contrast as a function wavelength. 
We followed a similar approach to the one defined in Section \ref{sec:analysis}: we fix the orbital parameters to the joint best value found in Section \ref{sec:analysis} and the spot position and radius to the values from the white-light analysis, and let the transit depth, the limb-darkening coefficient of the linear law and the spot contrast vary as a function of wavelength in each light curve. 
The resulting contrasts are presented in Figure \ref{fig:contrast}. 

Using stellar models to model the contributions of the spot and the stellar surface to the observed contrasts (see Appendix \ref{sec:spot-analysis} 
for details), we were able to derive spot temperatures for each of the events. 
For the cold spot observed in the April 2014 dataset, we find a temperature of $T_s = 5278 \pm 81$ K (or $\Delta T = 192 \pm 10$ K colder than the star). 
For the bright spot, we find a temperature of $T_s = 5588 \pm 92$ K (or $\Delta T = 137 \pm 10$ K hotter than the star). 
These models are also presented in Figure \ref{fig:contrast}.

\subsection{Stellar contamination}
\label{analysis:heter}


In addition to altering transit depths through active region crossing events, unocculted active regions can also alter observed transit depths through the transit light source effect.
We assessed the potential for stellar contamination of the transmission spectrum of WASP-19b through the transit light source effect using the method detailed by \citet{TLSE:2018}. 
The approach consists of placing constraints on allowed spot and faculae covering fractions using a set of rotating photosphere models and then translating the covering fractions into ranges of potential stellar contamination in the transmission spectrum. 

WASP-19 displays rotational variability at the 2\% level in the $V$ band (Section~\ref{sec:phot-mon}). 
Our transit observations show that both cool (spotted) and hot (facular) active regions are present in the photosphere (Section~\ref{sec:spot-crossing}). 
These active regions appear to be large, with radii of approximately 0.25 stellar radii, and display observed temperature contrasts of $\Delta T = -192 \pm 10$~K and $\Delta T = +137 \pm 10$~K, respectively. 
\citet{Mancini:2013} also observe WASP-19b transit with a spot-crossing event that displays a larger temperature contrast ($\Delta T = -680 \pm 50$~K) and smaller radius ($R_s=0.1651 \pm 0.0045$ stellar radii) than what we observe. 
Accordingly, we consider both smaller, high-contrast spots and larger, low-contrast spots in this analysis. 
In either case, as spots and faculae contribute approximately opposing variability signals, they can mask each other's presence in photometric monitoring \citep{TLSE:2018}. 
However, they also contribute nearly opposing stellar contamination signals and, therefore, the net contamination of the transmission spectrum arises from the relative overabundance of one signal. 
For this reason, we considered the effects of spots and faculae separately in this analysis, deriving end-member examples to examine the range of possible contamination signals. 
In total, we examined three possible sources for the rotational variability of WASP-19:  high-contrast spots, low-contrast spots, and faculae.

\begin{figure}
   \includegraphics[width=1\columnwidth]{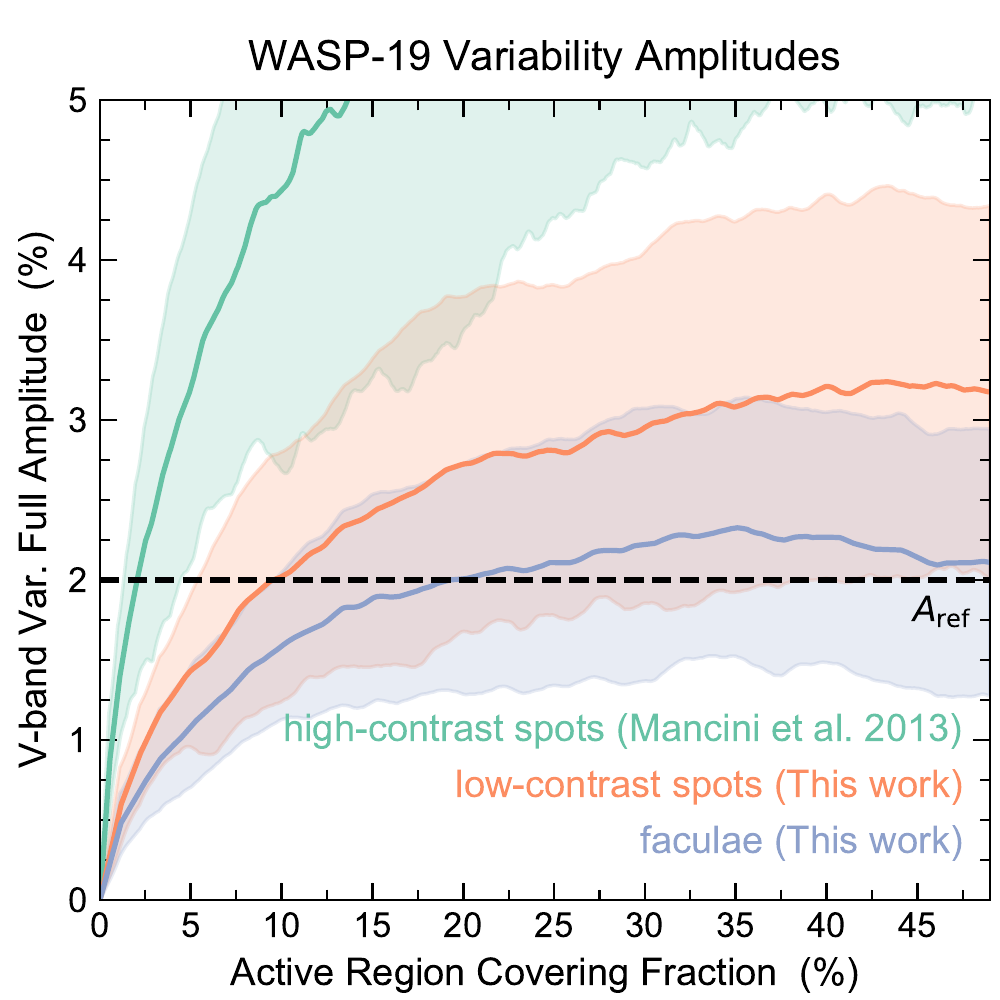}
\caption{Rotational variability amplitudes of WASP-19 in $V$ band as a function of active region covering fraction. Variability amplitudes grow with a square-root-like dependence on the active region covering fraction with a steeper dependence for higher-contrast heterogeneities. Results for high-contrast spots, low-contrast spots, and faculae models are shown in green, orange, and blue, respectively. Give the $1\sigma$ confidence intervals from the 100 model realizations, illustrated by shaded regions, the observed variability of WASP-19, $A_\textnormal{ref}=2\%$, is consistent with a wide range of active region covering fractions.}
    \label{fig:variability-amplitudes}
\end{figure}

For each case, we constructed 100 model photospheres, using PHOENIX model stellar spectra \citep{PHOENIX} of different temperatures to simulate contributions from the immaculate photosphere, cool spots, and hot faculae ignoring, for simplification, the impact 
of the position of the heterogeneities in the disk on the contrasts \citep[which could, in principle, produce changes in the spot and facular 
contrasts on the star; see, e.g., ][]{Wang:1998}. 
Utilizing constraints from the crossing events analyzed in Section~\ref{sec:spot-crossing} and by \citet{Mancini:2013}, we fixed the stellar parameters to the values in Table~\ref{tab:stellar-het} and interpolated within the PHOENIX model grid to generate component spectra with these parameters (namely, $F_{\lambda, \textnormal{phot}}$, $F_{\lambda, \textnormal{spot}}$, and $F_{\lambda, \textnormal{fac}}$).
We successively added active regions at random coordinates drawn from a uniform distribution and recorded the corresponding rotational variability. 
We find the mean and 68\% confidence intervals for the high-contrast spots, low-contrast spots, and faculae covering fractions consistent with the observed variability to be $f_\textnormal{spot,1} = 2.0^{+2.4}_{-0.7} \%$, $f_\textnormal{spot,2} = 10^{+30}_{-5} \%$, and $f_\textnormal{fac} = 19^{+31}_{-10} \%$, respectively (Figure~\ref{fig:variability-amplitudes}). 
The similarity of the estimates for the low-contrast spots and faculae owes to their similar temperature contrasts. 
High-contrast spots, however, generate larger variabilities and therefore can reproduce the observed variability of WASP-19 with a smaller covering fraction.

\begin{table}
    \centering
    \footnotesize
    \caption{Parameters used in stellar heterogeneity analysis. $T_\textnormal{phot}$, $[\textnormal{Fe}/\textnormal{H}]$, and $\log g$ are adapted from \citet{Doyle:2013}, $T_\textnormal{spot,1}$ and $R_{1}$ from \citet{Mancini:2013}, and the remainder are derived from this analysis.}
    \label{tab:stellar-het}
    \begin{tabular}{lcc} 
        \hline
        \hline
Parameter                          & Description                    & Value \\
        \hline
$T_\textnormal{phot}$              & Photosphere temperature        & 5460~K  \\
$[\textnormal{Fe}/\textnormal{H}]$ & Metallicity                    & 0.14 \\
$\log g$                           & Surface gravity                & 4.37 \\
$T_\textnormal{spot,1}$            & High-contrast spot temperature & 4780~K \\
$T_\textnormal{spot,2}$            & Low-contrast spot temperature  & 5270~K \\
$T_\textnormal{fac}$               & Facula temperature             & 5600~K \\
$R_{1}$                            & Small active region radius     & 0.17~$R_\textnormal{star}$ \\
$R_{2}$                            & Nominal active region radius   & 0.21~$R_\textnormal{star}$ \\
$A_\textnormal{ref}$               & V-band variability amplitude   & 2\% \\
        \hline
    \end{tabular}
\end{table}

\begin{figure}
   \includegraphics[width=1\columnwidth]{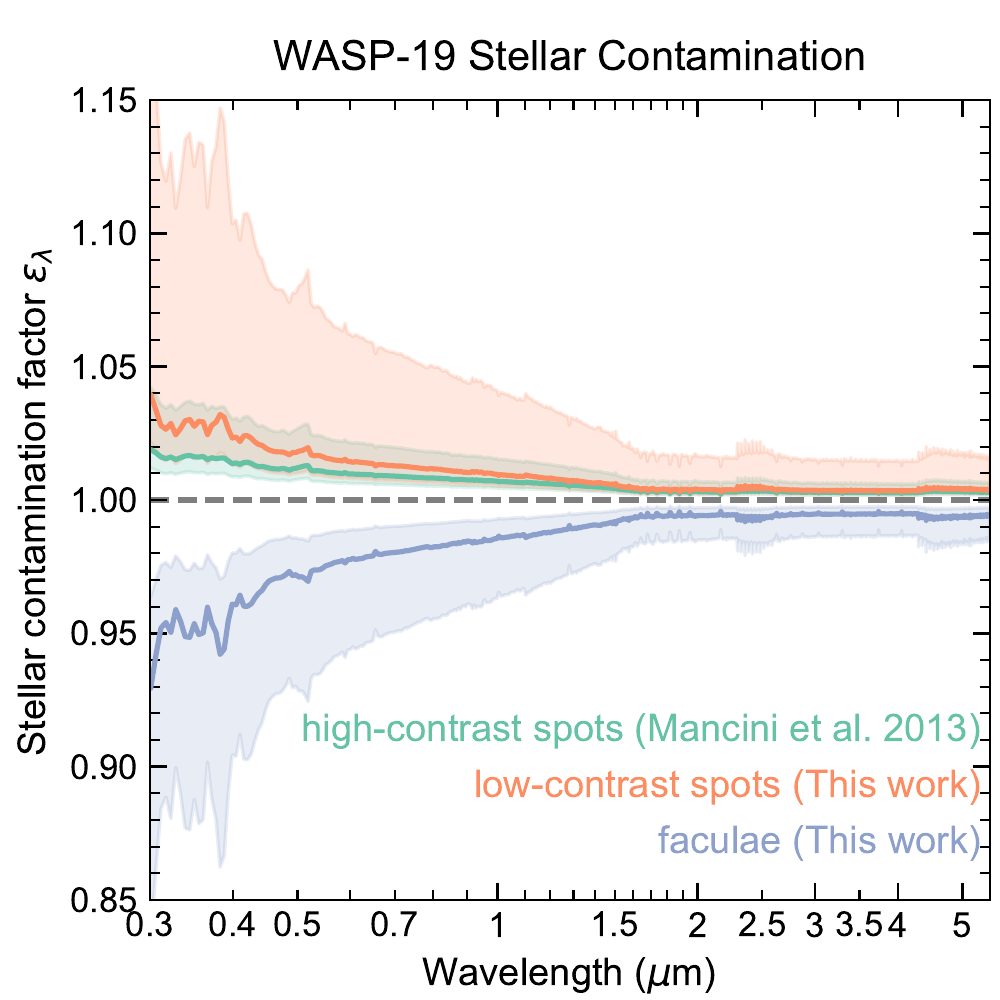}
\caption{Potential ranges of stellar contamination owing to unocculted active regions on WASP-19. The effects of unocculted high-contrast spots, low-contrast spots, and faculae are shown in green, orange, and blue, respectively. Lines show the expected contamination due to the mean estimate of the covering fraction of unocculted active regions, following our rotational photosphere modeling, and the shaded regions illustrate the range of stellar contamination corresponding to the $1\sigma$ confidence intervals on the active region covering fractions. At $1\sigma$ confidence, we estimate the effect of stellar contamination on our bluest measurements is no greater than $8.4\%$ of the transit depth or $1700$~ppm at 1-sigma.}
    \label{fig:CS}
\end{figure}

If present in the unocculted stellar disk, these active regions will not be directly evident in transit light curves, but will nonetheless impact transmission spectra by imparting a spectral difference between the disk-integrated stellar spectrum and the resolved region illuminating the exoplanet atmosphere during the transit (the actual light source used in transmission spectroscopy). Figure~\ref{fig:CS} shows the stellar contamination spectrum $\epsilon_{\lambda}$ produced by the spot and faculae, calculated as
\begin{equation}
\epsilon_{\lambda, \textnormal{het}} =
\frac{1}
     {1 - f_\textnormal{het}(1 - \frac{F_{\lambda, \textnormal{het}}}                                                         {F_{\lambda, \textnormal{phot}}})},
\label{eq:CS}
\end{equation}
in which the subscript ``het'' refers to either spots or faculae \citep[][Equation 2]{TLSE:2018}. 
The stellar contamination spectrum combines multiplicatively with the planetary transit depth to produce the observed transmission spectrum. 
In this case, low-contrast spots and faculae can have the largest effect on the observations; high-contrast spots have a comparably smaller effect given the lower allowed covering fractions. 
Low-contrast spots can increase transit depths at the shortest wavelengths of the IMACS observations (4500--4700~\r{A}) by $1.9\%$, considering the mean covering fraction estimate, or as much as $8.4\%$ for the $1\sigma$ upper estimate. 
Faculae, on the other hand, can decrease transit depths at these wavelengths by $3.1\%$ on average and as much as $7.7\%$ at $1\sigma$ confidence. 
Taking the transit depth of WASP-19b to be $20345$~ppm, these values correspond to absolute transit depth changes of $390$~ppm and $-630$~ppm for the mean spot and faculae coverages, respectively, and $1700$~ppm and $-1600$~ppm for their upper limits. 
Thus, at $1\sigma$ confidence, we estimate that stellar contamination can alter the transit depth of our bluest data by at most $1700$~ppm. 

Interestingly, as outlined in our analysis in Section \ref{sec:transpec}, we do see changes in the transit depths between different epochs on the order of these effects. Even more, the slope observed in the 17/02/11 dataset can, in fact, be explained as arising from stellar contamination from unocculted (low-contrast, cold) spots with a spot coverage within 1-sigma of what is predicted in Figure \ref{fig:CS} (see Section \ref{sec:discussion}). Another possibility to explain that slope is from \textit{occulted} bright spots: if not taken into account in the modelling, they could produce a slope similar to the one observed. However, there is no evidence in the white-light light curve for occulted spots (either bright or cold) on the 17/02/11 night. This would mean that any occulted bright spot is below the noise level in our white-light light curves, and thus we cannot confirm this possibility with our data.  

\section{Discussion}
\label{sec:discussion}
\subsection{Interpreting the Magellan/IMACS optical transmission spectra}
\begin{figure}
\includegraphics[height=0.6\columnwidth]{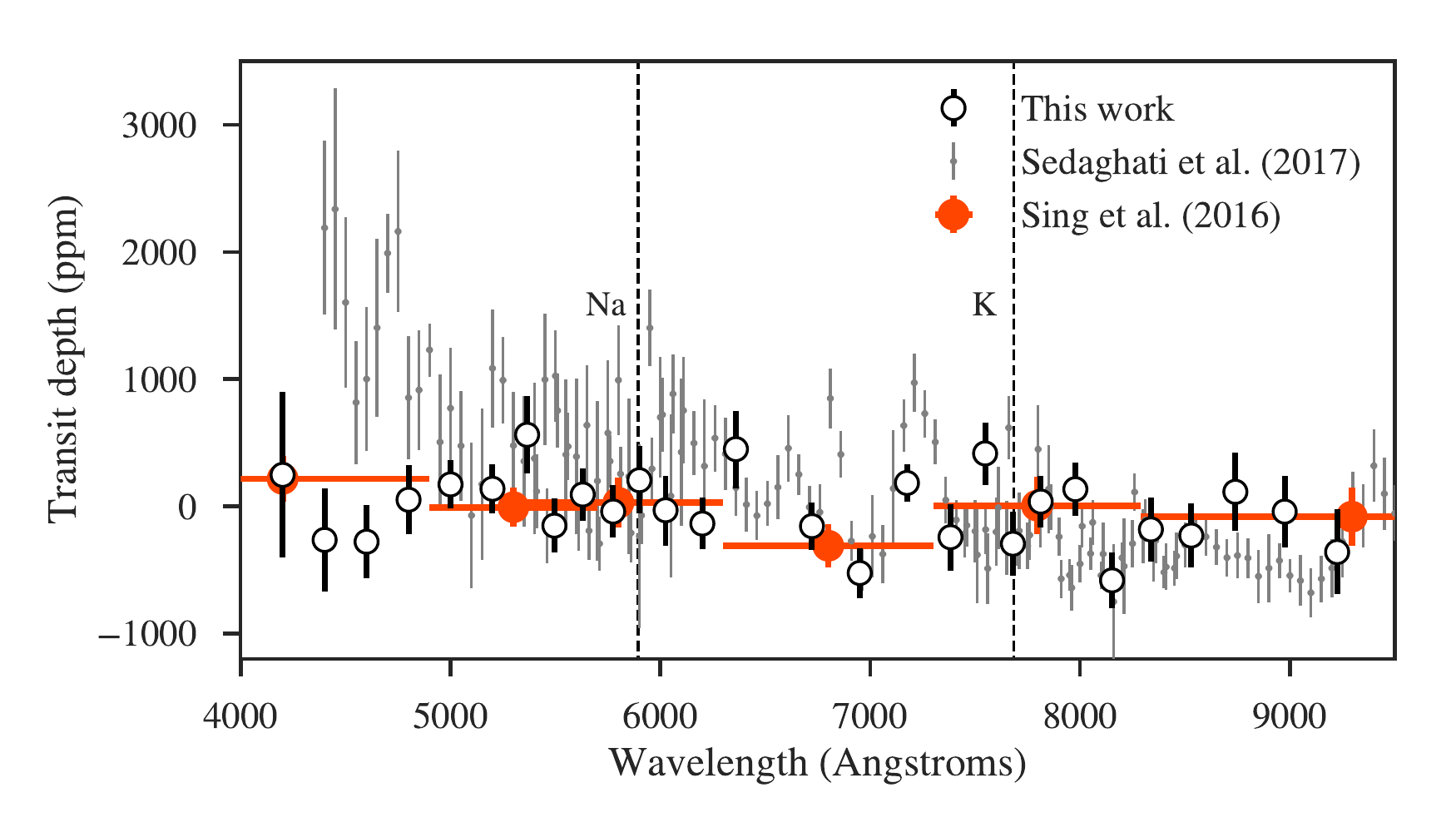}
    \caption{Combined transmission spectrum (white datapoints with errorbars) of WASP-19b from our Magellan/IMACS observations compared to the datasets in the same wavelength range presented in \protect\citet[][red points with errorbars]{Sing:2016} and in \protect\citet[][grey points with errorbars]{Sedaghati:2017}. The dashed lines indicate the position of the Na and K lines, which we do not detect in our data.}
    \label{fig:transpec-comparison}
\end{figure}

As presented in Section~\ref{sec:transpec} and Figure~\ref{fig:transpec}, our combined optical transmission spectrum from our Magellan/IMACS observations lacks spectral features at the couple of hundred parts per million level. 
This is in striking contrast with the transmission spectrum presented in \cite{Sedaghati:2017}, which shows a 
strong scattering slope and some features interpreted as TiO in that work. 
In Figure \ref{fig:transpec-comparison} we compare our mean-subtracted spectrum with those obtained by both \cite{Sedaghati:2017} and \cite{Sing:2016}; note the caveat, however, that the data of \cite{Sedaghati:2017} for wavelengths $\lesssim 6,000$~{\AA}, between $6,000-8,000$~{\AA}, and $\gtrsim 8,000$~{\AA} were taken in different epochs (and, therefore, possibly under different levels of stellar activity). 
It is interesting to see that the precision of the Magellan/IMACS spectrum is similar to that of the HST spectrum \citep{Sing:2016}, although with better wavelength resolution. 
However, the data from \cite{Sedaghati:2017} attains similar precision to ours but at an apparently better resolution; it is unclear exactly why this is so, although it might be due to our different binning strategies\footnote{For example, in \cite{FORSW19:2015}, where an early study of one of the WASP-19b datasets was presented, the authors 
binned the data in 200-\r{A} bins, with separations between the central 
wavelengths of each bin of 100 \r{A}. In principle, in that case 
each datapoint does not provide independent information on any possible features 
to be observed in the transmission spectrum, and for this reason we avoid performing 
such ``overlapping" binning of the data.
Of course, this gives rise to larger 
errorbars, but it avoids having correlations between adjacent wavelength bins 
in the transmission spectrum.}.

Figure \ref{fig:transpec-comparison} shows that the scattering slope bluewards of $\lesssim 5500$ \r{A} observed in \cite{Sedaghati:2017} is not evident in either our combined transmission spectrum or that of \cite{Sing:2016}. 
Interestingly, however, our February 2017 observations do show a decrease on the same order of magnitude as a function of wavelength of the transit depth, which could be due to stellar heterogeneities in the star (see Section~\ref{sec:170211}). 
It is thus possible that the observed slope in the data of \cite{Sedaghati:2017} is actually produced by starspots/faculae. 
This highlights the power of repeatability: 
especially in ground-based optical observations, repeated observations in the same wavelength range (and, if possible, also with different instruments/setups) are mandatory not only to increase the precision of the measurements but also their \textit{accuracy}, which in turn allows us to confirm that the features observed in transmission come indeed from planetary phenomena and not from systematic effects, either of stellar or instrumental nature. 
In addition to the absence of the scattering slope, we also do not observe clear signatures of sodium and potassium. 
Regarding TiO, by simply inspecting Figure \ref{fig:transpec-comparison} we can observe that the only possible sign of TiO absorption is with our datapoints at 7,000 \r{A} and 7,200 \r{A}. 
If there is a TiO feature, it is definitely smaller than (approximately half) the one presented in that work.

\subsubsection{Quantifying possible spectral signatures with atmospheric retrievals}
\label{sec:retrieval_framework}


To provide a quantitative measurement on the absence or presence of features, we performed an atmospheric retrieval on our data, simultaneously modeling the atmosphere of the planet and the spectral signal of any heterogeneities present in the star. 
For the former, we used the semi-analytical formalism for transmission spectroscopy on an isothermal, isobaric atmosphere presented in \cite{HK:2017}. 
For the latter, we used the formalism of \cite{Rackham:2017,TLSE:2018}, 
already presented in Section \ref{analysis:heter}. 

A detailed overview of the implementation of our retrieval framework can be found in Appendix \ref{sec:retrieval}. 
In summary, as with our spot-crossing analysis detailed in Section \ref{sec:spot-crossing}, we implement this framework using PyMultiNest \citep{MultiNest,PyMultiNest}, which allows us to compute bayesian evidences for the different possible models, allowing us to (1) quantitatively measure the evidence for different features in the atmosphere of the planet and (2) incorporate our ignorance on a ``best model" for the transmission spectra in the posterior distribution of the retrieved atmospheric parameters \citep[via Bayesian Model Averaging; ][]{BMA:2015}. 
Our full retrieval has initially 3 parameters of stellar origin: a temperature of the occulted (by the planet) stellar 
surface $T_\textnormal{occ}$, a temperature of the heterogeneous, unocculted surface of the star $T_\textnormal{het}$, and a fraction $f_\textnormal{het}$, which defines the fraction of the projected stellar disk covered by $T_\textnormal{het}$ (see Section \ref{analysis:heter}). 
The planetary atmosphere is defined by $5 + n$ parameters: $f$, a factor that scales the derived planetary radius in Section \ref{sec:white-light} in order to find the reference radius $R_0=fR_p$ at which the atmosphere is optically thick, $P_0$ the pressure at which the atmosphere is optically thick (which we could also interpret as a cloud-top pressure), two parameters that define our haze prescription \citep[the same as the one used in][see the Appendix for details]{Sedaghati:2017}, the atmospheric temperature $T$, and $n$ mixing ratios of the different elements considered in this work (Na, K, TiO, H$_2$O, CO$_2$, CO and CH$_4$). 
Using this framework, we now set to analyze the individual and combined transmission spectra of WASP-19b obtained with Magellan/IMACS.

\subsubsection{Analysis of the individual Magellan/IMACS transmission spectra}
\label{sec:170211}
We first analyzed each Magellan/IMACS optical transmission spectrum obtained 
in different epochs separately in order to test them individually for possible 
stellar contamination and/or planetary spectral features using our retrieval framework. 
$n_\textnormal{live}=2000$ points were used to explore the parameter 
space, and models were computed with and without heterogeneities as well as 
with and without hazes. For these retrievals, only Na, K, TiO and H$_2$O 
were considered as possible opacity sources in the planetary atmosphere, and 
all combinations of those elements were tried in order to see which model 
was preferred by the data. A simple flat line (i.e., a constant transit depth 
as a function of wavelength, which would be the expected signature for high 
altitude clouds at the resolutions and precision probed by our data) was also considered, as well as spectra purely dominated by stellar 
contamination (i.e., a flat planetary transmission spectrum multiplied 
by equation (\ref{eq:CS})). 

As qualitatively inspected in past sections, every dataset but the 17/02/11 dataset was consistent with models that neglected stellar contamination. 
Of those datasets, every dataset but the one obtained on 14/03/22 has the simple flat line as the model with the largest evidence (i.e., as the ``best-fit"), with all the other models being indistinguishable from it (i.e., all the atmospheric retrieval models have $\Delta \ln Z < 2$ with respect to that model). 
For the 14/03/22 dataset, there is a slight preference for models incorporating TiO ($\Delta \ln Z = 2.4$ in favor of the best TiO model compared to a flat line). 
However, the retrieved temperature for those models was unphysical (1,300 K) given WASP-19b's large equilibrium temperature, and thus we attribute this preference to statistical fluctuations rather than to a physical reason. 
Given our data, Na and K appear to not be present in the individual Magellan/IMACS optical spectra.

\begin{figure}
\includegraphics[height=0.525\columnwidth]{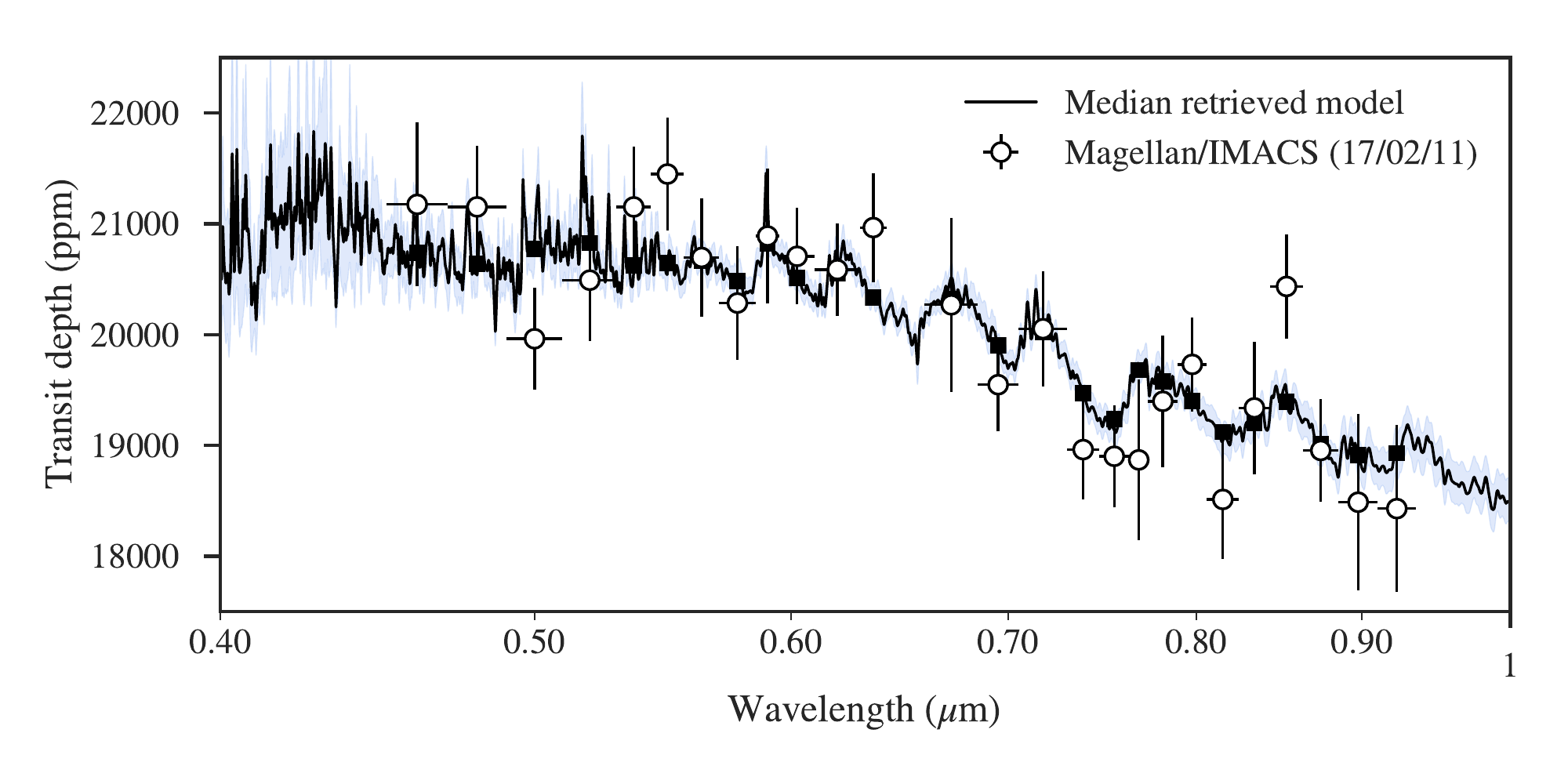}
    \caption{Transmission spectrum of our 17/02/11 night along with our best-fit 
    retrieval for that night, which considers a stellar contamination component, 
    along with an atmospheric planetary spectrum dominated by TiO, clouds and no 
    hazes.}
    \label{fig:transpec-170211}
\end{figure}

The 17/02/11 dataset is a very interesting one as the best-fit model, which is shown in Figure \ref{fig:transpec-170211}, includes stellar contamination, clouds \textit{and} TiO, but no hazes in the planetary atmosphere. 
This model is clearly preferred over a simple flat line ($\Delta \ln Z = 18.4$) and is slightly preferred over models that do not include a stellar contamination component with or without hazes and with or without TiO ($\Delta \ln Z > 2.7$). 
The model is, however, indistinguishable over a spectrum dominated purely by stellar contamination ($\Delta \ln Z = 1.3$). 
The caveat with this latter model is that in order to reproduce the observed transmission spectrum it requires that both the occulted and non-occulted parts of the photosphere have temperatures $<$ 4,500 K. On the other hand, the best-fit model shown in Figure \ref{fig:transpec-170211} has parameters whose posterior distributions are clearly bi-modal. 
One mode covers the same parameters as the just-described stellar contamination model: 
occulted and non-occulted temperatures having temperatures $<$ 4,500 K and, because those can match the data perfectly well without the contribution from a planetary atmosphere, low TiO volume mixing ratios ($\log_{10}\textnormal{TiO}<10^{-10}$). 
The other mode, however, has a complex structure. For the stellar contamination parameters, it gives
$f_\textnormal{het}=0.58^{+0.24}_{-0.21}$, 
$T_\textnormal{het}=5402^{+198}_{-230}$ K and 
$T_\textnormal{occ}=5780^{+148}_{-215}$ K. 
In this case, the stellar surface is consistent with WASP-19's effective temperature and, thus, it considers the scenario in which the average temperature of the occulted stellar photosphere by the planet is hotter than the average disk temperature.
Here, $58^{+24}_{-21}$\% of the stellar surface has a temperature $T_\textnormal{het}$ which is consistent with WASP-19's effective temperature. 
Interestingly, the temperature of the occulted part of the star by the planet in this case is, in fact, consistent within the errors with the bright spot temperature derived in Section \ref{sec:spot-crossing}; 
the areal covering fraction by this temperature of $1 - f_\textnormal{het} = 42^{+24}_{-21}$ \% is, in fact, also consistent within 1-sigma with the amount of variability observed in WASP-19b, according to our analysis in Section \ref{analysis:heter} (see Figure \ref{fig:contrast}). 
As for the parameters on the planetary atmosphere in this latter mode, in this retrieval the planetary temperature is $T=1951^{+542}_{-564}$ K --- very uncertain but consistent with WASP-19b's equilibrium temperature --- and 
the log-mixing ratio of TiO is $\log_{10} \textnormal{TiO} = -6.50^{+2.62}_{-3.26}$; 
uncertain but consistent with a solar-composition atmosphere. 

It is intriguing that the only dataset in which TiO is apparently detected is the 
one dataset in which there is a clear stellar contamination component, 
which was detected thanks to the large wavelength coverage of Magellan/IMACS. 
It might be that our TiO detection is, in fact, a spurious detection due to 
an oversimplification of our modelling of the stellar component present 
in the observed transmission spectrum and/or occulted spots in our data 
\cite[e.g., like suggested in][]{Oshagh:2014}. We tried a more complex parameterization of the 
stellar surface in which we consider both spots and faculae with different
disk-averaged and transit-chord covering fractions, as detailed by \cite{Zhang:2018}, but this only added complexity to 
the model without a further improvement of the fit: this more 
complex model has, in fact, an evidence $\Delta \ln Z = 7$ \textit{smaller} than 
the simpler model mentioned in the previous paragraph. It might be that there are 
stellar effects mimicking the TiO in this dataset that we are not considering (e.g., 
\textit{distribution} of spots with different sizes and temperatures, including spots 
with temperatures cooler than 4,500 K that could mimic atmospheric absorption by TiO in 
the transmission spectrum); a thorough search for 
such effects is, however, beyond the scope of the present study.

\subsubsection{A featureless combined optical transmission spectrum from Magellan/IMACS}

We now turn to the study of the combined optical transmission spectrum already presented at 
the beginning of this section. We ran the same retrievals as those for the 
individual transmission spectra just discussed in order to search for possible features 
in this more precise combined dataset. A featureless optical transmission spectrum (i.e., a 
constant transit depth as a function of wavelength) is the favored model given the data, 
with a log-evidence of $\ln Z = -200.6$; all the models including any one opacity source 
have larger evidences. In fact, the nominal model found by \cite{Sedaghati:2017} in their 
VLT/FORS2 dataset (which includes a haze, water, Na and TiO) has a log-evidence of 
$\ln Z = -204$ which means that, given our data, this latter model is 30 times less likely than the 
simpler flat transmission spectrum model in the optical. In summary, we find no evidence in our data 
for hazes and/or TiO absorption, and thus we conclude that, given our data, the transmission spectrum 
in the optical wavelength ranges probed by Magellan/IMACS is flat and consistent with high altitude clouds.
This conclusion is in agreement with predictions from the analysis of the water feature of \cite{Iyer:2016}, the interpretation of \cite{Barstow:2017}, and the lower-resolution HST/STIS optical transmission spectrum of \cite{Huitson:2013}. 

\subsection{The complete optical-to-IR transmission spectrum of WASP-19b}
\begin{figure*}
   \includegraphics[height=0.8\columnwidth]{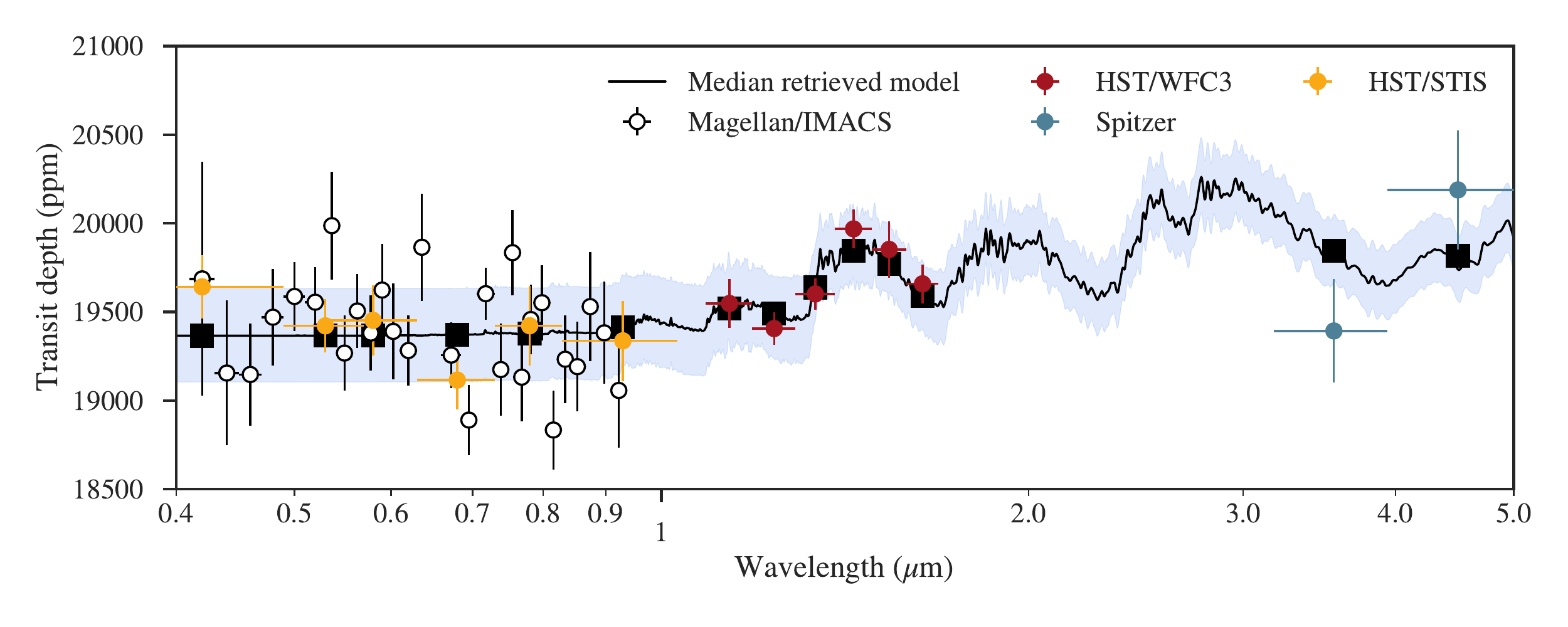}
    \caption{Complete optical to infrared transmission spectrum along with our ``best-fit" semi-analytical retrieval (which includes water absorption and a cloud deck, black solid line with blue bands which denote the 1-sigma contours) obtained by combining our Magellan/IMACS data (white datapoints) along with the available HST/STIS (yellow points with errorbars) and HST/WFC3 (red points with errorbars) data 
    from \protect\cite{Huitson:2013}, and the re-analyzed Spitzer (blue points with errorbars) data from \protect\cite{Sing:2016}. Black squares show the model binned to the HST and Spitzer data (Magellan/IMACS binning to the data is similar to the size of the datapoints, and it is thus not shown).}
    \label{fig:transpec_retrieval}
\end{figure*}

We now join our optical transmission spectrum from Magellan/IMACS with high-precision space-based observations to see what further constraints we can place on the interpretation of the complete optical-to-IR transmission spectrum of WASP-19b.
In addition to our optical observations, we consider the data obtained with HST in the optical (through HST/STIS) and IR (through HST/WFC3) by \cite{Huitson:2013} and \cite{Sing:2016} and the re-analyzed 3.6 and 4.5 $\mu$m Spitzer data for WASP-19b presented in \cite{Sing:2016}. We chose to use the data 
published in \cite{Sing:2016} and, e.g., not the analysis of the HST/WFC3 dataset presented in \cite{Mandel:2013}, as the data in \cite{Sing:2016} for HST/WFC3, HST/STIS and Spitzer has been 
analyzed uniformly. We perform the same retrievals as in the previous subsection but now including the most likely infrared absorbers (H$_2$O, CO, CH$_4$ and CO$_2$) and three additional parameters in order to account for three possible offsets in transit depth: 
one for our optical observations, one for the optical HST/STIS observations and one for the IR observations with HST/WFC3. 
These offsets are fitted in our retrieval in order to take into account possible overall mismatches due to stellar activity, in light of our discussion in Section \ref{sec:phot-mon} regarding the corrections made to the HST observations. 
To apply them, we mean-subtract those datasets, add the white-light transit depth $(R_p/R_*)^2$ calculated with the planetary and stellar radii value found in Section \ref{sec:white-light}, and consider constant offsets with gaussian priors centered in zero with wide standard deviations of 1000 ppm between the datasets. 
No offset is applied to the Spitzer data, as the impact of stellar activity at those wavelengths is expected to be negligible given the precision of the data and our analysis in Section \ref{analysis:heter}. 
Impacts of using different orbital parameters for the transit fit between these datasets and the Spitzer one are assumed to be negligible, as the ones used in \cite{Huitson:2013} and the ones used in this work have only small differences, with any impact on the transit depths being well within the Spitzer errorbars. 

There are many tens of retrieval models that are statistically indistinguishable between one another 
($\Delta \ln Z < 2$), almost all of which containing water as a common opacity source, no hazes, and 
no stellar heterogeneity component. 
The simplest model, and the one with the lowest evidence is a transmission spectrum purely dominated by clouds, no haze, no stellar contamination component and water absorption ($\ln Z = -304.5$), which is the retrieval presented in Figure \ref{fig:transpec_retrieval}. 
It is interesting to note that this model is statistically indistinguishable from a flat line ($\Delta \ln Z = 1.2$ in favor of the retrieval with water absorption), which highlights the fact that more data are needed in order to claim a strong water absorption detection in the infrared given the unknown transit depth offset between 
the optical and infrared data and the relatively large errorbars of the HST/WFC3 measurements. 

\begin{figure*}
   \includegraphics[height=0.7\columnwidth]{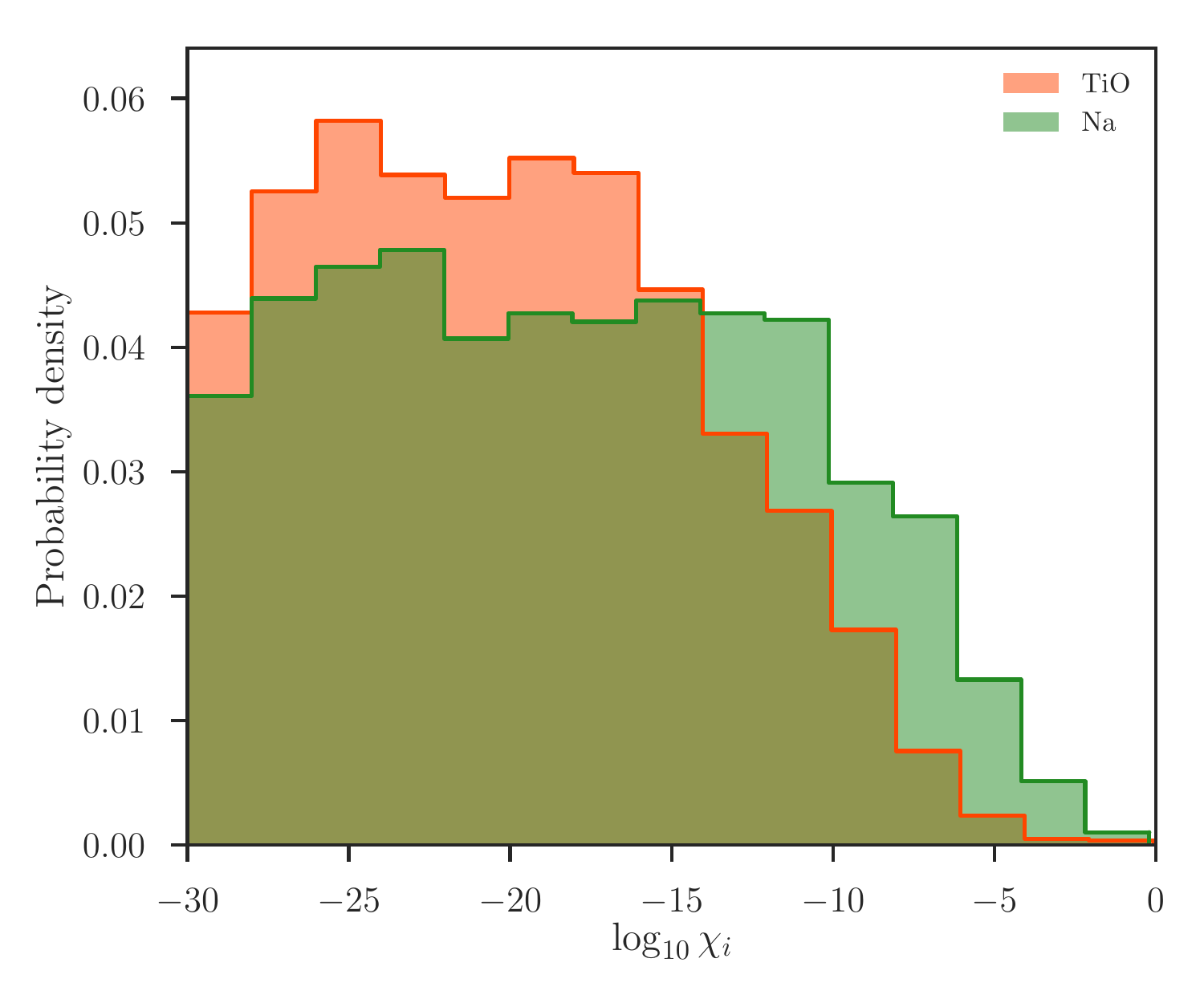}
   \includegraphics[height=0.7\columnwidth]{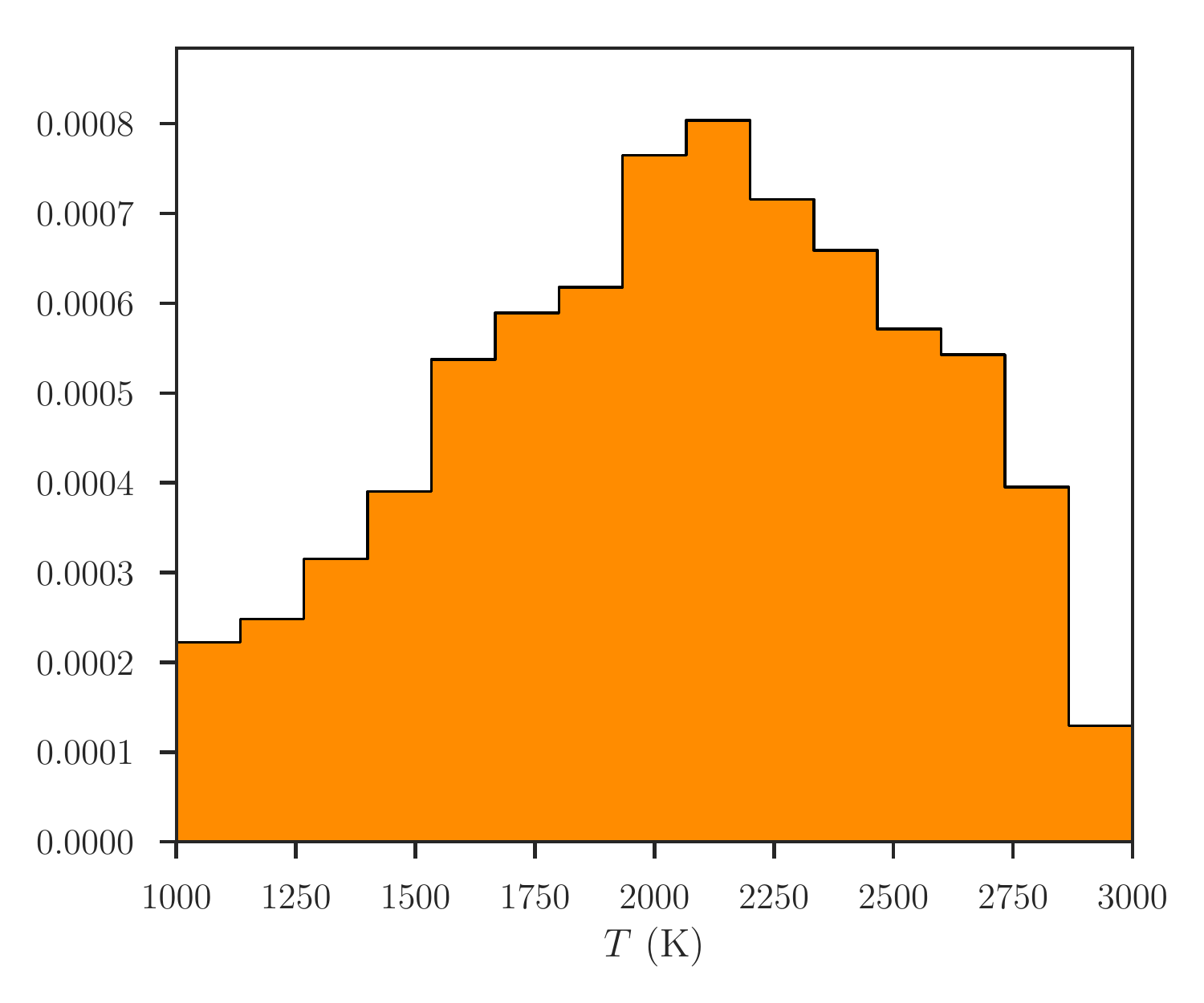}
   \includegraphics[height=1.6\columnwidth]{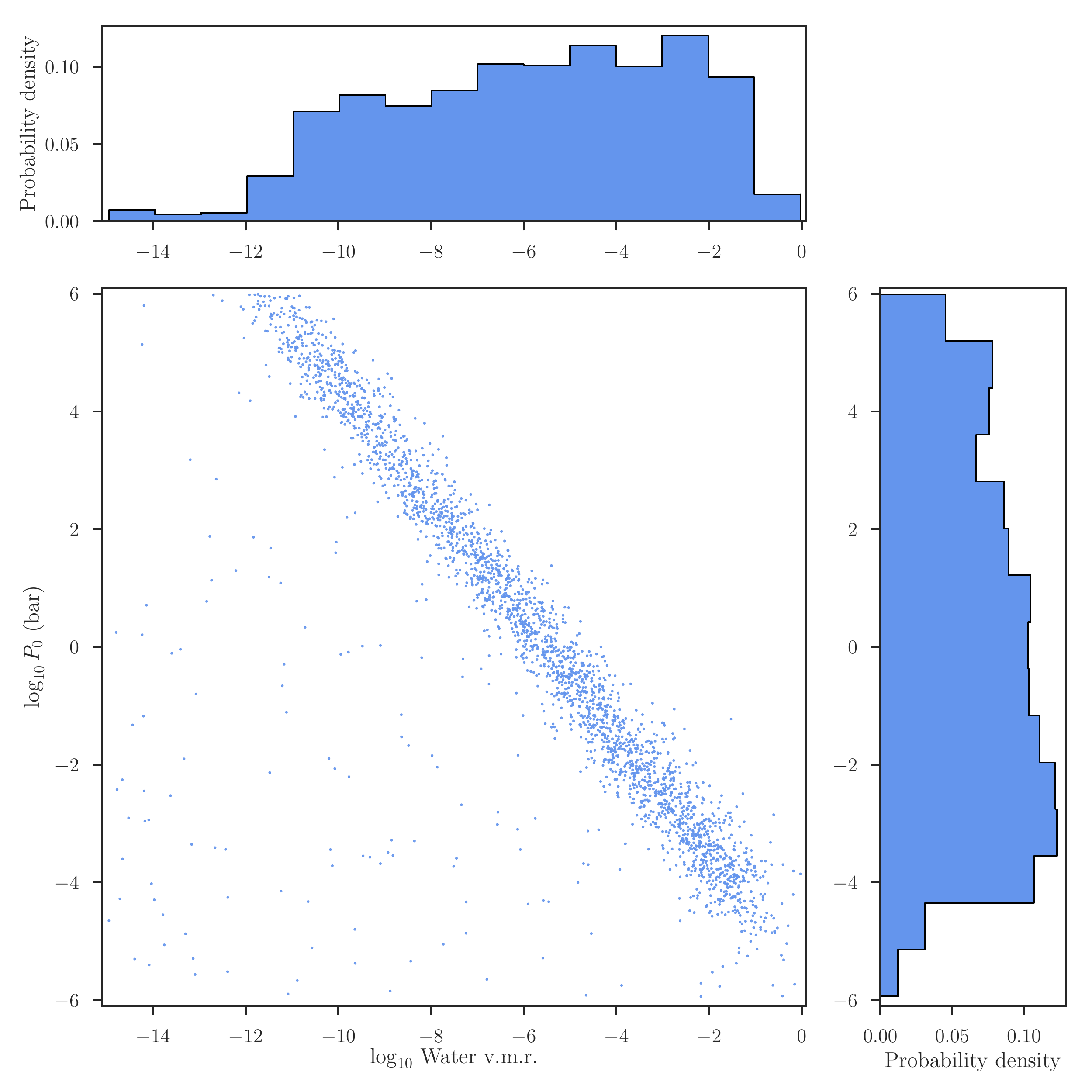}
    \caption{\textbf{(Top left)} Posterior distributions of the (log-10) volume mixing ratios, $\chi_i$ of TiO and Na, \textbf{(top right)} atmospheric temperature and \textbf{(bottom)} log-10 volume mixing ratio of water and \textbf{the pressure $P_0$} for WASP-19b obtained through our bayesian model averaging scheme over our model 
    retrievals (see text).}
    \label{fig:posteriors-bma}
\end{figure*}

In order to find the posterior distributions for the parameters of interest in our atmospheric retrievals, we perform bayesian model averaging \citep{BMA:2015} using the posterior distribution of all of our retrievals and our estimated evidences, assuming each model is equiprobable a-priori.
In this way, we get the marginalized (over our models) posterior distributions for all the parameters in our retrievals.

Figure \ref{fig:posteriors-bma} presents the most important parameters for our discussion. 
As can be seen, we obtain an uncertain atmospheric temperature of $2048^{+490}_{-540}$ K, whose distribution is peaked around the expected one from WASP-19b's equilibrium temperature of 2100~K (calculated using perfect day-night heat redistribution).
We also obtain an uncertain log-pressure (in bars) for the optically thick part of the atmosphere of $\log_{10} P_0 = -0.50^{+3.79}_{-2.81}$. 
The posterior distributions of both parameters are very interesting. 
On one hand, the retrieved terminator temperature is consistent with WASP-19b's day-side temperature estimated by \cite{Wong:2016} to be of $2372 \pm 60$ K. 
The retrieved distribution of the pressure $P_0$, on the other hand, is also interesting as it seems to be peaked towards small values (around 1 mbar), which we can interpret as the data being most consistent with high altitude clouds being present in WASP-19b. 
This interpretation is consistent with the one given by \cite{Barstow:2017}, who also retrieve high altitude clouds for WASP-19b with a different retrieval technique. 

We also obtain a very uncertain log-water volume mixing ratio of 
$\log \chi_{\textnormal{H}_2\textnormal{O}} = -5.83^{+3.30}_{-4.37}$. 
The mixing ratio of water under chemical equilibrium at 2048 K (our median retrieved atmospheric temperature) is $7.8 \times 10^{-4}$ at pressures of 0.3 bar (our median retrieved cloud-top pressure).
This is well within our posterior distribution and hence makes our retrieved water abundance consistent with that from a solar composition atmosphere. 
The weak constraint on the water mixing ratio from our atmospheric retrieval is most likely due to three effects: (1) as already discussed, the uncertain offset due to stellar activity between the infrared and optical observations; 
(2) the fact that the water signature in the infrared is not very strong, as already discussed in the previous paragraph; and 
(3) the correlation between $R_0$, $P_0$ and the mixing ratios \cite[see, e.g.,][and references therein]{HK:2017}. 
More data in this wavelength regime (or at any other water band in the infrared) might help put 
a tighter constraint on the water abundance in WASP-19b's terminator region. 

For TiO, we obtain a log-volume mixing ratio of $\log \chi_{\textnormal{TiO}} = -20.34^{+6.83}_{-6.19}$, which is also very uncertain, but with a distribution that clearly rules out the abundance expected in a solar-composition atmosphere. 
In this case, the expected abundance of TiO is $2.2 \times 10^{-7}$ at the same pressures and 
temperatures as before. 
The probability given our data that the mixing ratio has this or larger values is $\mathbb{P}(\chi_{\textnormal{TiO}} \geq 2.2\times 10^{-7}|D) = 1\%$.
Thus, in the framework of our semi-analytical retrievals, solar or super-solar TiO abundances in the terminator region of WASP-19b are very unlikely. 

For Na, a similar scenario arises: 
the expected solar composition abundance is $5.4\times 10^{-6}$. 
The corresponding probability of the mixing ratio having this or larger values is 
$\mathbb{P}(\chi_{\textnormal{Na}} \geq 5.4\times 10^{-6}|D) = 2\%$, which also suggests that solar and super-solar Na abundances in WASP-19b are very unlikely. 

Additionally, as a final note on the retrieved parameters, we find that hazes are not clearly preferred over simpler, non-hazy models.

Overall, our retrieved water abundance seems to be consistent with that of \cite{Sedaghati:2017}, 
but the abundance of TiO, the abundance of Na, and the presence of a haze are not. 
The latter discrepancies result from the fact that we do not detect their spectroscopic signatures in 
the optical in our combined Magellan/IMACS data, which is interesting and 
could perhaps be highlighting the possibility of stellar 
activity mimicking some or all of those features, as discussed in Section \ref{sec:170211}. The exact cause of the discrepancy between our 
combined transmission spectrum and that of \cite{Sedaghati:2017}, however, remains unknown.

\subsection{Giant spots on WASP-19}
We now discuss what may be gleaned regarding the spots on WASP-19 from our observations. 
In Section \ref{sec:spot-crossing} we reported sizes and temperatures for our observed spot-crossing events on the April 2014 and 2017 nights, with the former having a spot size of $R_s = 0.21 \pm 0.01$ stellar radii and a temperature $\Delta T = 192\pm 10$ K colder than the stellar photosphere, and the latter having a size of $R_s = 0.25^{+0.05}_{-0.04}$ stellar radii, and a temperature 
$\Delta T = 137 \pm 10$ K \textit{hotter} than the stellar photosphere. These sizes are very large, especially compared with the Sun, whose rarest, largest spots can reach $\sim 0.1R_*$ (see, e.g., sunspot AR 2192, which is the largest in the last 22-year solar cycle). 

Spots this size are, however, apparently not very uncommon on WASP-19.
Spots of sizes similar to the ones we obtain in this work were observed both in 2010 \cite[][who report a spot with $R_s=0.2635 \pm 0.0022$ stellar radii from two observations of the same spot]{Tregloan-Reed:2013} and 2012 \cite[][who report two spots, one with a size of $R_s=0.117 \pm 0.0125$ stellar radii, and one with a size of $R_s=0.1651 \pm 0.0045$ stellar radii]{Mancini:2013}. 
Interestingly, however, the \textit{temperature} of the detected spots apparently varies greatly. 
\cite{Mancini:2013}, for example, report a temperature $\Delta T = 683 \pm 103$ K cooler than the stellar photosphere for the largest spot observed in that work ($R_s=0.1651 \pm 0.0045$ stellar radii). 
This is significantly cooler than the $\Delta T = 192\pm 10$ K we observe for our cool, $R_s = 0.21 \pm 0.01$ stellar radii spot in 2014. 
This is very interesting because, on average, spots producing larger contrasts (i.e., having colder temperatures than the surrounding stellar photosphere) tend to be larger in our own Sun \citep{Wesolowski:2008}, which is the opposite to what we see between the spots in this work and the one in \cite{Mancini:2013}. 
In fact, if we compare the broadband sizes and contrasts quoted in the works of \cite{Tregloan-Reed:2013} and \cite{Mancini:2013} with ours, it appears the two largest starspots \citep[one of which is ours, and the other one from][]{Tregloan-Reed:2013} have the smaller contrasts, while the two smaller starspots 
\citep[both detected by][]{Mancini:2013} have the largest contrasts. 

There are various possibilities for explaining this behavior. 
One of them is that the relationship between size and contrast for the Sun found by \cite{Wesolowski:2008} has a large dispersion.
However, if we use the same relation for WASP-19 we cannot explain the observed variation: 
for large spot areas, the dispersion in contrast seems to be of order $\sim 0.05$, while the variation in contrast between our spots is of order $\sim 0.20$. 
Considering this observation only, the simplest explanation is that the size-contrast relation for our Sun does not apply for more active stars like WASP-19. 
On the other hand, derived spot contrasts could also be affected by unocculted stellar spots, leading us to make direct comparisons between contrasts obtained at different epochs very difficult. 
This effect has already been pointed out for measurements of absolute transit depths in transit light curves in \cite{TLSE:2018} and many other works, but here we argue that it could also impact  measurements of starspot contrasts, which depend on the relative dimming and brightening of the transit light curve during starspot crossing events\footnote{However, the spot size, which mostly depends on the start and end of this dimming event, is not impacted by this effect.}. 
In addition, it could also be that we are actually observing a collection of smaller spots within an active region and not giant spots themselves \cite[see, e.g., the discussion in ][]{firstfac:2016}; 
however, this hypothesis can only be tested with observations of spot-crossing events with better time resolution and photometric precision.

\section{Conclusions}
\label{sec:conclusions}
In this work, we have presented six optical transmission spectra of WASP-19b obtained in different 
epochs with the IMACS multi-object spectrograph mounted at the Magellan Baade 6.5m telescope, 
three covering the 4100-9350 \AA\ range, two covering the 4500-9350 \AA\ range and one covering 
the 4100-5560 \AA\ range. All but one spectrum covering the 4500-9350 \AA\ range are consistent 
with being flat, in striking contrast with the previous work of \cite{Sedaghati:2017}, who report 
a large blue optical slope and signatures of TiO, Na and H$_2$O in their optical transmission spectra. 
One of our transmission spectra, however, shows a clear decrease in transit depth towards 
longer wavelengths, which we interpret as arising mainly from stellar contamination. This is the only 
dataset in which, apart from the stellar contamination signature, TiO is tentatively detected, which 
is intriguing as it is also the dataset that shows the clearest signatures of being impacted by 
stellar contamination. Whether there is a causal relation between the two (either the TiO feature 
being from stellar origin or stellar activity inducing TiO features in WASP-19b's terminator region) 
is unknown. We combine our optical measurements along with previously published HST and Spitzer data 
in order to interpret WASP-19b's panchromatic transmission spectrum and use a semi-analytical retrieval 
approach to find a water abundance consistent with that from a solar composition atmosphere, along with 
sub-solar TiO and Na abundances in WASP-19b's terminator region.

In addition to our transmission spectroscopy study, we also study spot-crossing events that occurred 
in two out of the six observed transits, one of which we interpret as being caused by the crossing 
of a \textit{bright} spot, making this the first unambiguously detected bright spot-crossing 
event from an exoplanet transit light curve \cite[evidence for a bright spot-crossing event was 
first observed in WASP-52 by][]{firstfac:2016}. Both spots are very large ($\sim 0.2R_*$); however, 
this is apparently not uncommon on WASP-19b, especially given that spot-crossing events are more likely 
to be detected for these kind of giant spots. Using the wavelength-dependent light curves, we 
were able to constrain the temperatures of the spots using spectra from ATLAS model atmospheres \citep{ATLAS} to model the fluxes of both the stars and the 
spots. For the bright spot, we find it to be 
$\Delta T = 137 \pm 10$ K hotter than the stellar photosphere, while for the cold spot we find 
it to be $\Delta T = 192 \pm 10$ K cooler than the stellar photosphere. We use this information in 
order to give plausible scenarios for stellar contamination in WASP-19b's transmission spectrum 
using the work of \cite{TLSE:2018}, which in our case was fundamental to study in order to have 
a complete interpretation of the features observed by this and previous works on the transmission 
spectrum of WASP-19b.

Given our analysis and discussion, we conclude that more data are needed in the infrared in order 
to secure the HST/WFC3 water feature reported in \cite{Huitson:2013}. Such data will be fundamental 
to understand whether WASP-19b's metal content, as well as to know if the atmosphere has a low or high C/O ratio, which in turn could give 
clues as to the possible formation pathways that led to WASP-19b present-day spectra \citep{oberg:2011,madhusudhan:2012,mordasini:2016,Espinoza:2017}.

\section*{Acknowledgements}
We would like to thank an anonymous referee for suggestions and comments that 
improved this manuscript. N.E. would like to thank J.\ Araya, J.\ Bravo, P.\ Jones, M.\ Martinez, V.\ Meri\~no, M.\ Navarrete, H.\ Nu\~nez, A.\ Pasten and G.\ Prieto for numerous discussions (and subsequent experiments) done at Las Campanas Observatory (LCO) which were fundamental to better understand the many sources of systematics in the instrument, and all the staff at LCO for their invaluable support which made this work possible. N.E.\ would also like to thank R. 
McDonald for sharing the Na and K cross-section calculation method, and 
K.\ Heng for fruitful discussions regarding the semi-analytical formalism 
for transmission spectroscopy. N.E. would like to thank the Gruber Foundation for its generous support to this research. N.E.\ and A.J.\ acknowledge support from the Ministry for the Economy, Development and Tourism Programa Iniciativa Cient\'ifica Milenio through grant 
IC 120009, awarded to the Millennium Institute of Astrophysics.  A.J.\ acknowledges support by Fondecyt grant 1171208 and by CATA-Basal 
(PB06, CONICYT). N.E.\ acknowledges support from Financiamiento Basal PFB06 and from the CONICYT-PCHA/Doctorado Nacional graduate fellowship. 
B.R. acknowledges support from the National Science Foundation Graduate Research Fellowship Program under Grant No. DGE-1143953. 
D.A. acknowledges support from the Max Planck Institute for Astronomy, Heidelberg, for a sabbatical visit. We 
thank the anonymous referee for useful comments that improved this manuscript. The results reported herein benefited from collaborations and/or information exchange within NASA's Nexus for Exoplanet System Science (NExSS) research coordination network sponsored by NASA's Science Mission Directorate. This work has made use of the VALD database, operated at Uppsala University, the Institute of Astronomy RAS in Moscow, and the University of Vienna. This paper includes data gathered with the 6.5 meter Magellan Telescopes located at Las Campanas Observatory (LCO), Chile.




\bibliographystyle{mnras}
\bibliography{paperbib} 

\appendix
\section{Photometric monitoring of WASP-19}
\label{sec:phot-mon-an}
\subsection{2014 season}
\label{sec:phot-mon-2014}
\begin{figure}
   \includegraphics[width=\columnwidth]{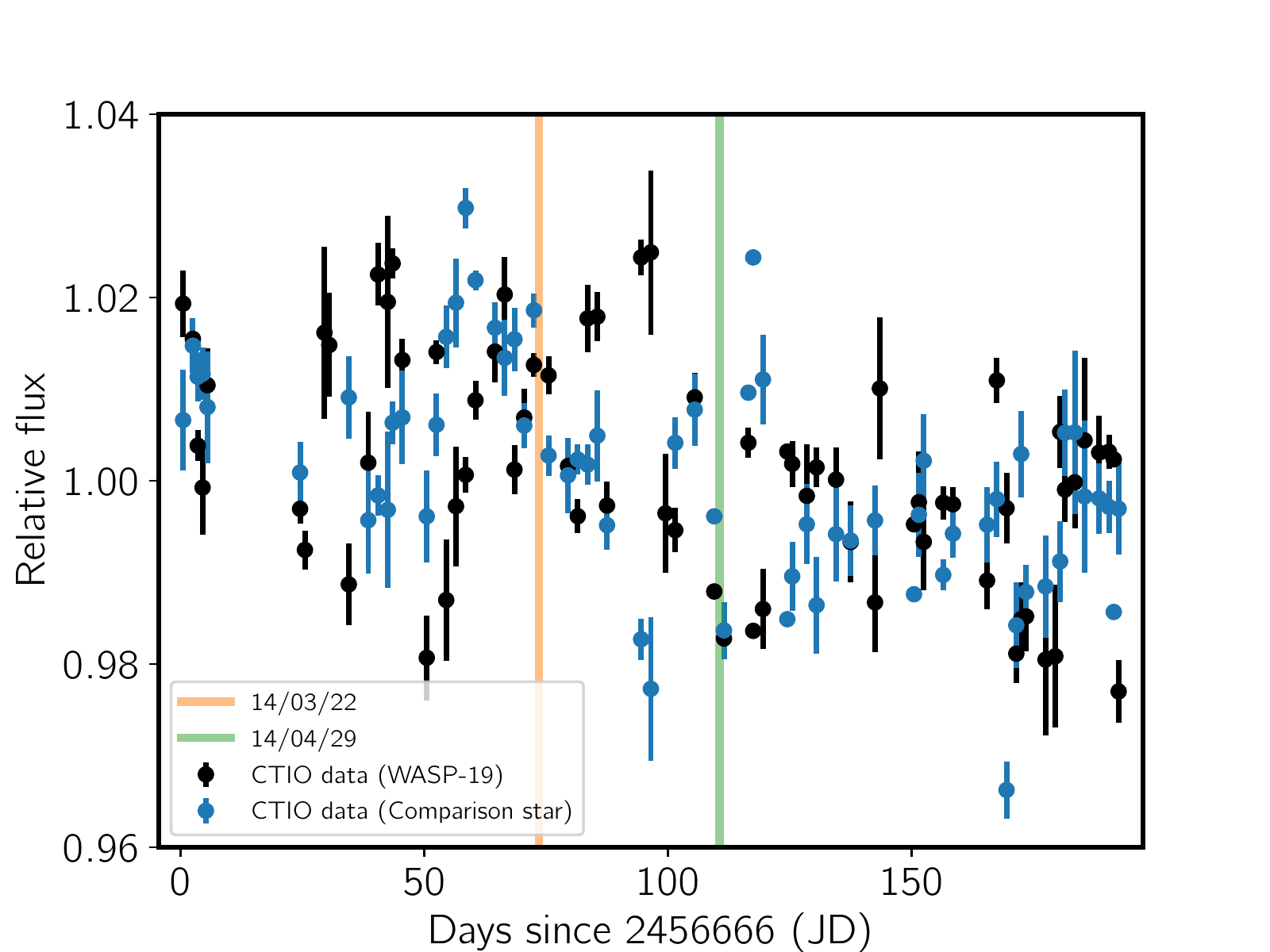}
    \caption{Photometric monitoring obtained during the 2014 season for WASP-19 from CTIO. The same photometry is shown
             for a comparison star. Note how the photometry is reliable only at the $\sim 1\%$ level, and displays
             night-to-night systematics visible in both WASP-19 and the comparison star.}
    \label{fig:CTIO-phot}
\end{figure}

For the 2014 season, WASP-19 was monitored with the SMARTS 1.3m telescope at CTIO with a $V$ filter. This project was a continuation of the CTIO photometric monitoring of WASP-19 presented in the 
work of \cite{Huitson:2013}. The data were downloaded from the NOAO archive\footnote{\url{http://archive.noao.edu/}}, 
and reduced with a standard photometric pipeline that performs aperture photometry on the stars using the \texttt{Astropy} package \citep{Astropy:2013}. The five stars closest in brightness to WASP-19 
were used to obtain the relative photometry of WASP-19. The resulting photometry is shown in Figure~\ref{fig:CTIO-phot} along 
with that obtained for a comparison star close in brightness to WASP-19, in order to illustrate the long-term photometric precision and stability achieved by these observations.

Although the formal precision ($\sim 0.3\%$) should have allowed us to 
detect variability at levels $\lesssim 1$\%, the photometry for both WASP-19 and the comparison star vary significantly 
more than the formal errors computed assuming photon noise (up to 2\% in some portions of the light curve) and thus are of insufficient quality to observe the rotational modulations of the star. The conditions in this 
season were similar to those of the 2013 season shown by \cite{Huitson:2013}, in which precisions of the same order of magnitude 
were reported. Because of this, we decided to re-reduce the data published in that work which is also available from the 
NOAO archive in order to test for any problems with our reductions and/or methods. Doing this we recovered the same variability 
and precision reported in that work. However, we observed the same qualitative behaviour as that shown in Figure~\ref{fig:CTIO-phot}: 
the variability is present not only in WASP-19 but also in the comparison stars. This implies that the observed variability 
reported in \cite{Huitson:2013} is not entirely of astrophysical origin but, rather, mostly due to a strong instrumental component 
like that observed in Figure~\ref{fig:CTIO-phot}. This large instrumental scatter is most likely due to 
the fact that the nightly observations were obtained by placing WASP-19 and the comparison stars at different positions in the detector every night. 
Different methods were implemented in order to try improve the photometry, but none of them gave good results.

We describe this dataset in detail because 
\cite{Huitson:2013} corrected the {\em HST} transmission spectrum assuming the 
observed variability was {\em only} astrophysical in origin, i.e., due to the rotational modulation caused by stellar spots. However, as we 
have shown, the variability in the CTIO dataset is mostly spurious.
Therefore, the correction to the transmission spectrum done in \cite{Huitson:2013} is not appropriate. This correction mostly adds an overall shift in the transit depth, because the effect as a function of 
wavelength is smaller than the optical errorbars published in that work and it has a negligible impact on the infrared. 
We discuss this in the following sections when deriving and comparing the optical transmission spectrum of WASP-19b.

\subsection{2017 season}
\label{sec:phot-mon-2017}
\begin{figure*}
   \includegraphics[width=2\columnwidth]{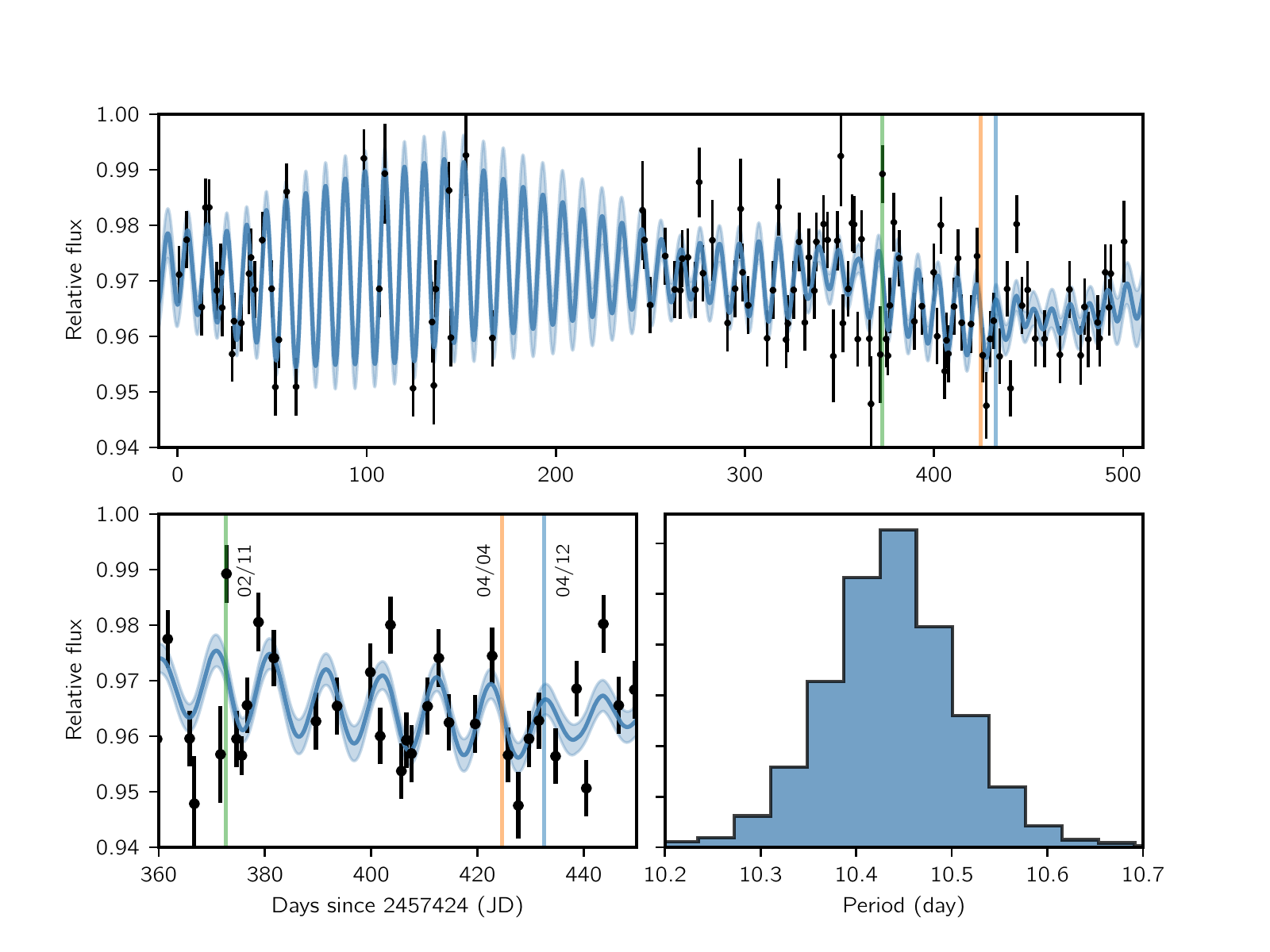}
    \caption{Photometric monitoring obtained during the 2017 season for WASP-19 from ASAS-SN. \textit{Upper panel}. Photometry 
             of WASP-19, along with vertical bands indicating the epochs at which the Magellan+IMACS observations were carried 
             out: green for the 17/02/11 observations, orange for the 17/04/04 observations and blue for the 17/04/12 
             observations. The blue curve that goes horizontally through the points is the best-fit gaussian process used to 
             both retrieve the rotation period of the star and to estimate the relative flux level of WASP-19 at those 
             different epochs (see text). \textit{Lower panel}. The left lower panel shows a close-up of the photometry and 
             best-fit gaussian process (see text) to around the epochs of the Magellan+IMACS observations. The right lower panel 
             shows the posterior distribution of the period estimated through our gaussian-process regression (see text).}
    \label{fig:ASASN-phot}
\end{figure*}

For the 2017 season, we used the photometry published by the ASAS-SN, whose precision suffices for the purpose of detecting stellar activity of 
relatively bright stars such as WASP-19. 
In Figure~\ref{fig:ASASN-phot} we show the photometry and we also indicate the epochs in which the observations with 
Magellan/IMACS where performed. In order to analyze this photometry, we fitted a Gaussian process regression using 
\texttt{celerite} \citep{celerite}. In particular, we use a kernel defined by
\begin{eqnarray*}
k (\tau) = \frac{B}{2+C}e^{-\tau/L}\left[\cos\left(\frac{2\pi\tau}{P_\textnormal{rot}}\right) + (1+C)\right],
\end{eqnarray*}
where $\tau = t_i - t_k$, with $i,k \in [1,2,...N]$, where $N$ is the number of datapoints, and $B$, $C$, $L$ and $P_\textnormal{rot}$ 
are the hyperparameters of the model, with the latter corresponding to the period of the quasi-periodic oscillations defined by 
this kernel. The usage of this kernel and its application to the recovery and modelling of stellar activity has been discussed already
in \cite{celerite}; we refer the reader to that work for the details. However, we briefy mention here that this kernel has the key 
properties of any kernel used to model stellar activity: it is quasi-periodic (due to the cosine term), it captures additional long-term trends in 
the data not captured by quasi-periodic modulations (due to the multiplicative exponential term), and it has the key property of being positive 
for all values of $\tau$.

In order to fit this Gaussian process to the data, we first obtained the maximum-likelihood estimates of the parameters through a simple optimization using \texttt{scipy}'s \texttt{optimize} function. With these best-fit parameters as input, we ran a 
Markov Chain Monte Carlo (MCMC) using the \texttt{emcee} ensemble sampler \citep{emcee}, by using 100 walkers around this 
best-fit value, and letting them explore the parameter space for 200 steps, with 100 steps as burn-in, and wide priors for the parameters being optimized. This resulted in 100,000 samples from the posterior density. The posterior probability 
distribution obtained for the rotation period ($P_\textnormal{rot}$) is shown on the lower-right panel of Figure~\ref{fig:ASASN-phot}. 
The posterior value is $P_\textnormal{rot} = 10.44\pm 0.07$ days, which is in excellent agreement with 
the $10.5$-day period reported in \cite{Hebb:2010} and in \cite{Huitson:2013}.

The flux obtained and modelled for WASP-19 with the ASAS-SN photometry is relative. However, it is desirable to find an estimate of 
the \textit{absolute} flux, taking into account 
the fact that the maximum brightness does not correspond to an unspotted disk in general. In order to take this into account, we obtained an estimate of the absolute flux level of the star in the observed band using the approximate method outlined 
in \cite{APZ:2012}, which divides the relative flux measurements by $F_* = F_\textnormal{max} + \sigma_F$, where $F_\textnormal{max}$ is the maximum observed brightness 
of the star and $\sigma_F$ is the standard deviation of the light curve. We used the retrieved Gaussian process conditional on our data in order to obtain $F_*$; this predicted photometry 
is shown in the upper and lower-left panels of Figure~\ref{fig:ASASN-phot}. Our Gaussian process regression returns an estimate for the absolute 
flux in the $V$-band for WASP-19 relative to an immaculate star of 
$0.972\pm 0.003$ for the 17/02/11 observations, of $0.964 \pm 0.003$ for the 17/04/04 observations and 
$0.966\pm 0.003$ for the 17/04/12 observations, i.e., a fairly similar flux level for our observations made in April, but a slightly brighter 
flux level on our February observations. It is interesting to note, however, that during the February observations, the ASAS-SN photometry seems to vary significantly more around the time of the observations.

\section{Updating the transit ephemerides for WASP-19b}
\begin{figure}
   \includegraphics[width=\columnwidth]{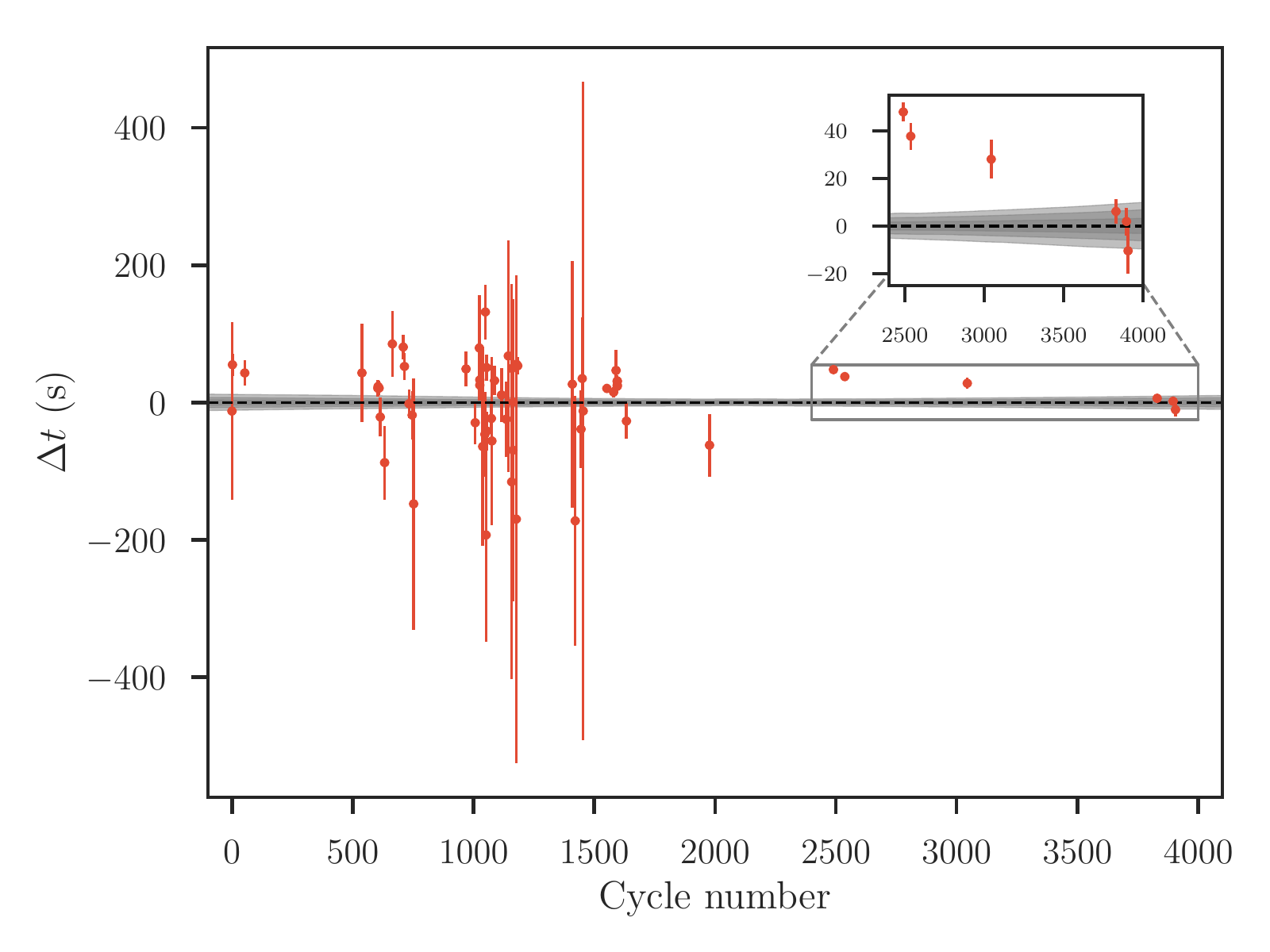}
    \caption{Residual timings (red points with errorbars) from our fit (grey bands depicting the 1,2 and 
    3-sigma credibility bands) to the transit ephemerides of WASP-19b. The inset shows a zoom to our 
    Magellan/IMACS datapoints, where it can be observed that the 2014 and 2015 observations (three leftmost datapoints 
    in the inset) are clear outliers of the fit.}
    \label{fig:ephem}
\end{figure}
\label{sec:ephem}
In order to update the ephemerides of WASP-19b using our measured transit mid-times 
we first transform our times in Table \ref{tab:system-params} which are in BJD UTC to BJD TDB by adding $32.184 +N$ seconds to our BJD UTC times, 
where $N$ is the number of leap seconds elapsed since 
1961\footnote{Obtained from the \texttt{TAI-UTC} column in \url{http://maia.usno.navy.mil/ser7/tai-utc.dat}}, 
which were 35 for our 2014-2015 season and 37 for the 2017 season \citep[note this 
implies we are omitting relativistic corrections, which account for as much as 
$\sim 1.6$ milliseconds --- much less than our best attained timing precision of 
4 seconds for the 14/03/22 observations. See ][]{Eastman:2010}. We couple those BJD 
TDB times with the ones in \cite{Mancini:2013}, and derive a new ephemerides (in BJD TDB 
units) of
\begin{eqnarray*}
t_0 = 2454775.337777(42) + 0.788839316(17)E,
\end{eqnarray*}
where, as in \cite{Mancini:2013}, $E$ is the number of cycles elapsed since 
the first transit observed by \cite{Hebb:2010}. This new ephemerides is both 
consistent with \textit{and} more precise than the one derived in \cite{Mancini:2013}. 
We observe, however, that there is a systematic difference of around 40 
seconds between the timings on our 2014-2015 observations and the 2017 
observations (see Figure \ref{fig:ephem}). Given that these have errors between 4--10 seconds, the 
timing difference seems to be significant. A full analysis of the entire 
transit timing dataset, which is beyond the scope of this work, could elucidate
what could be producing these changes in transit time, which is also seen in 
previous epochs \citep[but see][who conservatively choose to not attribute those to real 
variations but to underestimations of the timing errorbars]{Mancini:2013}.

\section{Spot-crossing events analysis}
\label{sec:spot-analysis}
The analysis of the spot-crossing events described in this work were separated in two 
steps. On the first step, we analyzed the band-integrated (``white-light'') light curves 
in order to obtain the size of the spots, and then this parameter was used in order to 
fit the wavelength-dependent light curves with which we were able to obtain wavelength-dependent 
constrasts for the spots.

For the analysis of the band-integrated transit distortions, we gave as inputs to \texttt{spotnest} the systematics-corrected light curves along with the priors on the transit parameters obtained in Section \ref{sec:analysis}, and let $n_\textnormal{live}=500$ points explore the parameter space, with which the log-evidences were calculated with errors below $0.1\%$ for spotless, one-spot and two-spot models, and with which we were able to generate $3500$ samples from the posterior distribution for each model. The results for our ``best" models (in terms of the calculated evidences) are presented in Figure \ref{fig:spot-wl}. 

For the April 2014 dataset, the spot models are favored with very high significance, as expected from the observed light curve distortions ($\Delta \ln Z = 163$ in favor of the one-spot model when compared to the spotless model which, in the case of both models being equally likely a priori, would imply that the model with spots is $(Z_\textnormal{spot})/(Z_\textnormal{no-spot})= 6 \times 10^{70}$ times 
more likely than the non-spotted model). However, the two-spot model is not favored by our data: the difference in log-evidence is actually higher for the one-spot model, with $\Delta \ln Z = 2$ in favor of this simpler model. The resulting one-spot model led to a very precise measurement on the position of the spot on the (projected) stellar surface of $(x,y)=(0.21^{+0.01}_{-0.01},0.68^{+0.04}_{-0.06})$, where $x$ goes in the direction almost perpendicular to the planet's motion and the center of the star is at $(0,0)$, with contrast of $F_\textnormal{spot}/F_\textnormal{star} = 0.84^{+0.01}_{-0.03}$, i.e., a dark spot, and a considerably large radius of $R_\textnormal{spot} = 0.21^{+0.01}_{-0.01}$ stellar radii. For the April 2017 dataset, the spot model is also favored with high significance, with a $\Delta \ln Z = 58$ between the spotless and one-spot models in favor of the latter. Again, the two-spot model is not favored by our data, with a $\Delta \ln Z = 2$ in favor of the simpler, one-spot model. The position of the spot is also identified with high precision to be at $(x,y)=(0.41^{+0.02}_{-0.02},0.43^{+0.08}_{-0.08})$ in the stellar surface, with a size of $0.25^{+0.05}_{-0.04}$ stellar radii. The most interesting feature of the April 2017 feature, however, is its contrast: the retrieved spot contrast is $F_\textnormal{spot}/F_\textnormal{star} = 1.16^{+0.05}_{-0.04}$, which implies the feature is actually a \textit{bright} spot. This makes the April 2017 event, thus, one of the first unambiguously detected bright spot features, i.e. a facular or plage region, on an exoplanet host star \citep[along with that of][ on WASP-52]{firstfac:2016}, and the first one on WASP-19.

In order to extract the spot contrasts as a function of wavelength, we performed the same analysis on the systematics-removed, wavelength-dependent light curves, following a similar approach to the one defined in Section \ref{sec:analysis}: we fix the orbital parameters to the joint best value found in Section \ref{sec:analysis}, the position of the spot on the stellar surface and the spot radius, and let the transit depth, the limb-darkening coefficient of the linear law and the spot contrast to vary as a function of wavelength on each light curve. Once all the contrasts were obtained for both datasets, we used ATLAS stellar model atmospheres \cite{ATLAS} in order to constrain the temperature of the spots using the wavelength-dependent spot contrasts, by assuming that the spots produce the same flux as a star with the same properties except for the temperature. In order to perform this fit, model spectra for stars from 4000 K to 6500 K were used, with gravities and metallicities similar to WASP-19 according to the works of \cite{Hellier:2011} and \cite{Doyle:2013} (i.e., $\log g=4.5$ and $[\textnormal{Fe}/\textnormal{H}]=0.1$). We interpolated these spectra in order to have a function that returned the shape of the spectrum for a given temperature. With this function, we fitted the contrasts by a two-parameter function $F(T_s)/F(T_{W19})$, where $F(T)$ is the (integrated, in the respective wavelength bins) flux at a given temperature, $T_s$ is the temperature of the spot and $T_{W19}$ is the temperature of WASP-19. For $T_s$ a uniform prior is used between $4,000$ K and $6,500$ K, while for $T_{W19}$ we put a normal prior with the temperature measured by \cite{Doyle:2013} for WASP-19, which is $5,460 \pm 90$ K. The fit then was free to explore the parameter space, with the constraint that for the cold spot $T_s<T_{W19}$ and for the bright spot $T_s>T_{W19}$. The resulting contrasts and their best-fit variations as a function of wavelength with this simple model are shown in Figure 
\ref{fig:contrast}.   

The fits to the observed contrasts are excellent. For the cold spot observed in the April 2014 dataset, we find a temperature of $T_s = 5278 \pm 81$ K (or $\Delta T = 192 \pm 10$ K colder 
than the star). The variation as a function of wavelength is highly significant in this dataset, with a $\Delta \ln Z = 20$ in favor of a variation in wavelength versus no variation (i.e., constant contrasts as a function of wavelength). For the bright spot, we find a temperature of $T_s = 5588 \pm 92$ K (or $\Delta T = 137 \pm 10$ K hotter than the star), and find moderate evidence in favor of a variation as a function of wavelength with a log-evidence difference of $\Delta \ln Z = 3$ when compared with a constant contrast as a function of wavelength, which in the equiprobable case, would imply the variation as a function of wavelength is 20 times more likely than a contrast which is constant as a function of wavelength. 

\section{Semi-analytical atmospheric retrievals}
\label{sec:retrieval}
Our implementation of the semi-analytical framework for transmission spectroscopy of 
\citep{HK:2017} follows the equation that describes the radius as a function of wavelength in 
transmission of a transiting exoplanet, $R(\lambda)$, which is given by
\begin{eqnarray*}
R(\lambda) = R_0 + H\left[\gamma + E_1 +  \ln\left(\tau_0\right) \right],
\end{eqnarray*}
where
\begin{eqnarray*}
E_1 = -\gamma - \ln \tau_0 + \int_0^{\tau_0} \frac{1-e^{-\tau}}{\tau}d\tau. 
\end{eqnarray*}
Here $H=k_BT/\mu g$ is the atmospheric scale-height given the mean molecular weight, $\mu$, the temperature, $T$, the planetary gravity, $g$, and Boltzmann's constant, 
$k_B = 1.38\times10^{-16}$ ergs/K, $\tau_0 = (P_0 \sigma/kT) \sqrt{2\pi HR_0}$ is 
a reference optical depth, that is defined by the (temperature and 
wavelength-dependent\footnote{Note that the cross-section is also pressure dependent. However, due to the low ranges of pressures usually probed in transmission, and the high expected temperature of WASP-19b, we omit the pressure-dependence of the cross-sections in this work.}) cross section $\sigma$, a reference radius $R_0$ and a reference pressure $P_0$. $\gamma = 0.57721$ is the Euler-Mascheroni constant. For the cross section, we use
\begin{eqnarray*}
\sigma(\lambda,T) = \sigma_\textnormal{haze} + \sum_{i=0}^N \chi_i \sigma_i(\lambda,T),
\end{eqnarray*}
where $\chi_i$ is the volume mixing ratio of species $i$ which has a cross-section 
$\sigma_i(\lambda,T)$ and where $\sigma_\textnormal{haze}(\lambda) = a\sigma_0(\lambda/\lambda_0)^{\gamma_\textnormal{haze}}$ is used to emulate 
hazes in the atmosphere; this is the same haze prescription used in \cite{MacDonald:2017}, with $\sigma_0 = 5.31 \times 10^{-27}$ cm$^2$, $\lambda_0 = 3500$ \r{A} and free parameters $a$ and ${\gamma_\textnormal{haze}}$. In this formalism, $R_0$ can be interpreted as a ``surface" where the atmosphere is optically thick (i.e., where the opacity $\tau \to \infty$), which can be produced either by the increased atomic/molecular absoportion deep in the atmosphere and/or 
to a cloud deck below which no radiation passes through; as such, clouds are naturally considered here. Similarly, $P_0$ can be interpreted as the 
pressure where this optically thick portion of the atmosphere lies, and so the transmission spectrum is actually generated 
by interaction of species at pressures $P>P_0$. We let both of these as free parameters in our retrievals; the pressure is explicitly fit, while for $R_0$ we 
fit for a factor, $f$, and set $R_0 = fR_p$, where $R_p$ is the 
planetary radius found in Section 
\ref{sec:white-light}. For any additional grey or close-to-grey opacity source, the parametrization 
of $\sigma_\textnormal{haze}(\lambda)$ covers various ranges of scattering sources, from 
grey opacities ($\gamma_\textnormal{Haze}\to 0$) to Rayleigh scattering-like 
opacities ($\gamma_\textnormal{Haze}\to -4$).

To model heterogeneities in the 
star we used the framework outlined in Section \ref{analysis:heter}, i.e., we 
use equation (\ref{eq:CS}) to model any stellar contamination simultaneously with 
our transmission spectrum retrieval. This is parametrized in our model by 
three terms: a temperature of the occulted (by the planet) stellar 
surface $T_\textnormal{occ}$, a temperature of the heterogeneous, unocculted 
surface of the star $T_\textnormal{het}$, and a fraction $f_\textnormal{het}$, 
which defines the fraction of the projected stellar disk covered by $T_\textnormal{het}$. 
The variation as a function of wavelength of 
those portions of the stellar surface is modelled with PHOENIX model atmospheres, 
as described in Section \ref{analysis:heter}.

We fix the planetary gravity to 1414 cm/s$^2$ which was the one derived in Section 
\ref{sec:white-light} and as possible opacity sources in the atmosphere of WASP-19b, 
we consider Na, K, TiO, H$_2$O, CO$_2$, CO and CH$_4$. All cross-sections but that of Na and K are calculated using \texttt{HELIOS-K} \citep{heliosk}. The 
line-lists for H$_2$O, CO$_2$ and CO are obtained from the HITEMP database 
\citep{hitemp}, while those of CH$_4$ are obtained from the ExoMol database \citep{exomol:2016,CH4:2013}. TiO is synthesized using the latest (2012 version) 
line lists calculated by B. Plez, obtained from the Vienna Atomic Line 
Database \citep[VALD;][]{VALD}. For Na and K, we use the analytical lorentzian-profiled doublet shapes used in 
\cite{MacDonald:2017}. All the cross sections are calculated at 1 mbar --- low 
enough as for pressure broadening effects to be negligible, effects that are small and that we would not be able to detect in this work. We consider the atmosphere 
to be dominated by molecular hydrogen, with a mixing ratio of H$_2$ of 0.83 
and He of 0.17 (which is based on chemical equilibrium calculations). 

As explained in Section \ref{sec:discussion}, our full retrieval has initially 3 parameters of stellar origin ($T_\textnormal{occ}$, $T_\textnormal{het}$, $f_\textnormal{het}$), and 
$5 + n$ parameters from the planetary atmosphere: $f$, $P_0$, the haze 
parameters $a$ and $\gamma_\textnormal{haze}$, the atmospheric temperature $T$ 
and $n$ mixing ratios of the different elements considered. As with our 
spot modelling, we explore the parameter space using MultiNest \citep{MultiNest} 
with the PyMultinest Python wrapper \citep{PyMultiNest}, which allows 
us to properly compare the probability of different models given the data. For 
the parameters of stellar origin we consider uniform priors between 
2,500 and 6,500 K for the temperatures and a uniform prior for the fraction 
$f_\textnormal{het}$ between 0 and 1. For the planetary atmosphere parameters 
we set a wide uniform prior between 0.8 and 1.2 for $f$, a wide log-prior 
between $10^{-6}$ and $10^6$ bars for $P_0$, a wide log-prior between $10^{-30}$ 
and $10^{30}$ for $a$, a wide uniform prior from -50 to 10 for 
${\gamma_\textnormal{haze}}$ and a uniform prior between 1,000 and 3,000 K for 
the atmospheric temperature $T$. For the volume mixing ratios of the elements 
under consideration we set log-uniform priors between $10^{-30}$ and 1.

\bsp    
\label{lastpage}
\end{document}